\documentclass[journal,draftclsnofoot,onecolumn,12pt]{IEEEtran}

\normalsize
\usepackage{mathtools}
\usepackage{ifpdf}
\usepackage{cite}
\usepackage{adjustbox}
\usepackage{algorithm}

\usepackage{amssymb}
\usepackage{amsmath}
\usepackage{amsthm}
\usepackage{algorithmic}
\usepackage{array}
\usepackage{epstopdf}

\usepackage{epsfig}
\usepackage{color}
\usepackage{tabularx}

\newtheorem{theorem}{Theorem}

\usepackage{graphicx}
\usepackage{caption}
\usepackage{subcaption}
\usepackage{mathtools}

\DeclarePairedDelimiter\floor{\lfloor}{\rfloor}

\begin{document}
\title{Construction of Protograph-Based Partially Doped Generalized LDPC Codes}
\author{Jaewha Kim, Jae-Won Kim, and Jong-Seon No
\thanks{J. Kim is with the Department
of Electrical and Computer Engineering, INMC, Seoul National University, Seoul 08826, Korea (e-mail: woghk9307@snu.ac.kr).}
\thanks{J.-W. Kim is with the Department of Electronic Engineering, Engineering Research Institute (ERI), Gyeongsang National University, South Korea (e-mail: jaewon07.kim@gnu.ac.kr).}
\thanks{H.-Y. Kwak is with the Department of Electrical Engineering, University of Ulsan, Ulsan 44610, South Korea.}
\thanks{J.-S. No with the Department
of Electrical and Computer Engineering, INMC, Seoul National University, Seoul 08826, Korea (e-mail: jsno@snu.ac.kr).}
\thanks{J.-W. Kim is the corresponding author.}
}
\markboth{}%
{Submitted paper}

\maketitle

\begin{abstract}
In this paper, we propose a new code design technique, called partial doping, for protograph-based generalized low-density parity-check (GLDPC) codes. 
While the conventional construction method of protograph-based GLDPC codes is to replace some single parity-check (SPC) nodes with generalized constraint (GC) nodes applying to multiple variable nodes (VNs) that are connected in the protograph, the proposed technique can select any VNs in the protograph to be protected by GC nodes. 
In other words, the partial doping technique facilitates finer tuning of doping, which in turn enables a sophisticated code optimization with higher degree of freedom. We construct the proposed partially doped GLDPC (PD-GLDPC) codes using the partial doping technique and optimize the PD-GLDPC codes by the protograph extrinsic information transfer (PEXIT) analysis. 
In addition, we propose a condition guaranteeing the linear minimum distance growth of the PD-GLDPC codes and use the condition for the optimization. 
Experimental results show that the optimized PD-GLDPC codes outperform the conventional GLDPC codes and have competitive performance compared to the state-of-the-art protograph-based LDPC codes without the error floor phenomenon over the binary erasure channel (BEC). 
\end{abstract}


\IEEEpeerreviewmaketitle

\section{Introduction}
\label{sec:introduction}
{L}{ow}-density parity-check (LDPC) codes, first introduced in~\cite{Gallager}, have received much attention due to their low decoding complexity and capacity approaching performance \cite{MacKay}. 
An LDPC code is defined over a bipartite graph consisting of variable nodes (VNs) and single parity-check (SPC) nodes. 
As a generalized class of LDPC codes, generalized LDPC (GLDPC) codes were introduced in~\cite{Tanner}, which are constructed by replacing some SPC nodes with generalized constraint (GC) nodes. 
GC nodes are defined by code constraints of a linear code with a larger minimum distance~\cite{doping}, which makes GLDPC codes have a larger minimum distance~\cite{dmin_GLDPC_IT}. In addition, GLDPC codes have several advantages over LDPC codes such as faster decoding convergence~\cite{fast_convergence} and a better asymptotic threshold at the cost of the additional decoding complexity and redundancy introduced by GC nodes~\cite{IT_GLDPC}. 
Many types of linear codes for GC nodes, also called as the component codes, are used in the GLDPC codes such as Hamming codes \cite{HammingGLDPC}, Hadamard codes \cite{HadamardGLDPC}, Bose–Chaudhuri–Hocquenghem (BCH) codes, and Reed-Solomon (RS) codes \cite{RSBCH_GLDPC}.
The research on GLDPC codes is extended to spatially coupled LDPC codes \cite{SC-LDPC1,SC-LDPC2,SC-LDPC3} and doubly GLDPC codes \cite{DGLDPC1,DGLDPC2,DGLDPC3}.
Moreover, some capacity approaching GLDPC codes were constructed using irregular random GLDPC codes \cite{IT_GLDPC,random_GLDPC}. \\
\indent {LDPC codes can be constructed from a small bipartite graph called protograph. Many researches on the protograph-based LDPC codes were previously carried out under various scenarios}~\cite{Proto_survey1,Proto_survey2,Proto_survey3}.
{Moreover, the protograph-based GLDPC codes were thoroughly studied in [21]-[24], but they mainly focused on the low-rate codes}~\cite{Proto_GLDPC1,Proto_GLDPC2,Proto_GLDPC3,Proto_GLDPC4}.
Protograph-based GLDPC codes can be constructed from a small protograph~\cite{Thorpe} using the so called doping technique~\cite{QC_GLDPC}. 
{Doping a GC node, defined by a $(\mu,\kappa)$ linear code of length $\mu$ and dimension $\kappa$, means the replacement of an SPC node by the GC node with $\mu-\kappa$ constraints, which causes a rate loss.
In the perspective of VNs, $\mu$ VNs are selected to be doped by a GC node, assuming that there are no parallel edges in the protograph. 
Thus, the smallest unit of doping, also called the doping granularity, is $\mu$
 for the conventional protograph doping technique. }
In other words, the conventional doping technique has two limitations: 1) the degree of the SPC node to be replaced should be $\mu$, which implies that the doping operation is dependent on the underlying protograph and the parameter $\mu$ of component codes and 2) one cannot choose a finer doping granularity less than $\mu$ and thus the code design cannot be sophisticated.
Due to the limited design flexibility, there has been little works on the well-designed optimization for protograph-based GLDPC codes especially for medium to high code rates. \\
\indent In this paper, we propose a new doping technique, called partial doping on the VNs, to minimize the doping granularity and enlarge the code design freedom.
In detail, the partial doping involves the following three steps: 1) A VN to be doped is selected in the protograph. 2) The Tanner graph is obtained by the lifting operation~\cite{Thorpe} from the protograph with a lifting factor $N$. 3)~Additional GC nodes are connected to the lifted $N$ VNs in the Tanner graph after lifting the protograph.
The main difference from the conventional protograph doping technique is that the partial doping operation is conducted on the Tanner graph instead of the protograph domain.
Thus, it is possible to partially dope on a single VN in the protograph and the doping granularity becomes one, which is also independent of $\mu$. 
In other words, the partial doping enables fine tuning of the code structure regardless of the underlying protograph and the parameter of component codes. 
Specifically, the selection of VNs to be protected by GC nodes and the rate loss can be adjusted in a more flexible manner.\\
\indent We denote the proposed protograph-based GLDPC codes constructed using the partial doping as partially doped GLDPC (PD-GLDPC) codes. 
The structural characteristics of the PD-GLDPC codes have several advantages.
First, the PD-GLDPC codes are structurally adequate to adopt the puncturing technique that compensates the rate-loss.
Since the partially doped VNs are highly and locally protected by GC nodes, the performance loss occurred by puncturing the doped VNs is relatively small while attaining the code rate gain.
Second, the asymptotic performance of the PD-GLDPC codes can be analyzed by the low-complexity extrinsic information transfer (EXIT) analysis. 
For the conventional protograph doped GLDPC codes~\cite{QC_GLDPC}, the exact EXIT analysis is provided in~\cite{exact_EXIT}, where the topology for the a priori and extrinsic mutual information of GC nodes is considered.
Since the cases of the topology grow exponentially with the component code length $\mu$, the computational complexity is too high to design a fast optimization algorithm. 
On the contrary, GC nodes in the PD-GLDPC codes can be analyzed by an average manner EXIT analysis in~\cite{EXIT_TIT} because GC nodes in the PD-GLDPC codes are incident to VNs lifted from a single VN in the protograph.
The a priori and extrinsic mutual information of GC nodes can be evaluated by a single value, which facilitates a fast optimization algorithm. 
Using this advantage, we propose an efficient optimization algorithm for the PD-GLDPC codes. \\
\indent In addition, we propose the condition guaranteeing the linear minimum distance growth of the PD-GLDPC codes. 
We analytically prove that the PD-GLDPC code ensembles satisfying the condition have the typical minimum distance and use this condition for the construction of the PD-GLDPC codes in this paper.
{Also, we propose novel methods to optimize the asymptotic performance, i.e., the threshold of the code ensemble, by using the protograph EXIT (PEXIT) analysis}~\cite{PEXIT} {and differential evolution}~\cite{Differential evolution} {targeting medium code rate $1/2$ and high code rate $2/3$}.
Thus, the optimized PD-GLDPC code ensembles are constructed while satisfying the typical minimum distance condition to have a minimum distance that grows linearly with the block length of the code.
Comparison of the PD-GLDPC codes is made with the existing state-of-the-art protograph LDPC codes and conventional GLDPC codes~\cite{random_GLDPC}.
Threshold analysis and shows that the optimized protograph-based PD-GLDPC codes outperform the well known GLDPC and protograph-based LDPC codes and have a competitive asymptotic performance compared to the optimized protograph-based LDPC codes. \\
\indent {To be specific, the optimized protograph PD-GLDPC codes from a random ensemble with a low doping ratio $0.02439$ achieves the coding gain $0.0079$ over the binary erasure channel (BEC) compared to the optimized GLDPC codes}~\cite{random_GLDPC}~{with a relatively higher doping ratio 0.4. In addition, the optimized protograph PD-GLDPC code by the differential evolution outperforms AR4JA codes}~\cite{div_proto}~{with coding gains $0.0477$ and $0.032$ for code rates $1/2$ and $2/3$, respectively. 
Also, the average VN degree of the optimized PD-GLDPC codes, are only $87.2\%$ and $80.5\%$ compared to the state-of-the-art protograph LDPC codes for code rates $1/2$ and $2/3$, respectively.}
Similarly, the frame error rate (FER) results show tangible gain in the waterfall performance compared to the existing protograph-based LDPC codes in~\cite{div_proto} without the error floor phenomenon up to FER $10^{-4}$. \\
\indent{We list the contributions of this paper as follows;
1) We propose a novel doping technique, where the constraints of GC nodes are applied to specific VNs lifted from single protograph node, i.e., partial doping on the VNs after lifting. 
2) We propose two design criteria for the optimization of the threshold of the PD-GLDPC codes: the EXIT analysis and the condition for the existence of the typical minimum distance.
3) {We propose the optimization method of the asymptotic performances for the PD-GLDPC codes using differential evolution.}
4) We show the finite length performance gain of the optimized PD-GLDPC codes over some well known LDPC and GLDPC codes. 
}\\
\indent The rest of the paper is organized as follows. 
In Section II, we introduce some preliminaries on the BEC and protograph-based GLDPC codes. 
Section III illustrates the proposed PD-GLDPC code structure and derives its PEXIT analysis and the condition for the typical minimum distance. In addition, the comparison of the proposed PD-GLDPC codes and protograph doped GLDPC codes is given. 
The optimization algorithms of PD-GLDPC codes are given in Section~IV. 
Section V shows the error correcting performance of the proposed codes over the BEC compared with other well known protograph-based LDPC codes.
Section VI concludes the paper with some discussion of the results. 

\vspace{10pt}
\section{Backgrounds}
\begin{table*}[]
\caption{{Main mathematical notations used in the paper.}}
\resizebox{\textwidth}{!}{%
\begin{tabular}{|l|l|}
\hline
\multicolumn{1}{|c|}{\textbf{Notation}}                  & \multicolumn{1}{c|}{\textbf{Explanation}}                                                                 \\ \hline\hline
${\bold{B}}_{n_c \times n_v}$                   & An $n_c \times n_v$ base matrix defining the protograph                                          \\ \hline
$V=\{v_1,{\cdots},v_{n_v}\}$                    & Set of VNs in a protograph                                                                       \\ \hline
$C=\{c_1,{\cdots},c_{n_c}\}$                    & Set of CNs in a protograph                                                                       \\ \hline
$E$                                             & Set of edges connecting $V$ and $C$                                                              \\ \hline
$deg(v_j)$ ($deg(c_i)$)& Number of edges incident to $v_j$ ($c_i$), i.e., variable (check) node degree of $v_j$ ($c_i$)  \\\hline
$N(c_i)$ ($N(v_j)$)   & Set of variable (check) nodes incident to $c_i$ ($v_j$), i.e., neighborhood of $c_i$ ($v_j$)     \\\hline
$I_{ch}(j)$                                     & Channel information of $v_j \in V$                                                               \\ \hline
$I_{EV}(i,j)$  $(I_{EC}(i,j))$                  & Extrinsic information sent from $v_j$ ($c_i$) to $c_i$ ($v_j$)                                   \\ \hline
$I_{AV}(i,j)$ $(I_{AC}(i,j))$                   & A priori mutual information of $v_j$ ($c_i$) sent from $c_i$ ($v_j$), where $c_i$ is an SPC node \\ \hline
$I_{AGC}(i)$ $(I_{EGC}(i,j))$                   & A priori (extrinsic) mutual information of a GC node $c_i$                                       \\ \hline
$I_{APP}(j)$                                    & A posteriori probability of $v_j$                                                                \\ \hline
$(\mu,\kappa)$ component code                   & Component code with codelength $\mu$ and information size $\kappa$                               \\ \hline
$\mathcal{X}$                                   & Index set of protograph VNs that are partially doped                                             \\ \hline
$N$                                             & Lifting factor                                                                                   \\ \hline
$\nu$                                           & Doping ratio                                                                                     \\ \hline
$I_{EV}^{(b_j)}(j)$                             & Extrinsic information from $v_j$ to $b_j, j\in \mathcal{X}$                                      \\ \hline
$I_{AGC}^{(b_j)}(j)$ $(I_{EGC}^{(b_j)}(j))$     & A priori (extrinsic) mutual information of $b_j, j\in \mathcal{X}$                                \\ \hline
$G_c$                                           & Optimized protograph of the LDPC code                                                            \\ \hline
$G_p$                                           & Initial irregular protograph that is used to construct the PD-GLDPC code                         \\ \hline
$\lambda_{G_c}(x)$ $(\rho_{G_c}(x))$            & VN (CN) degree distribution to construct $G_c$                                                   \\ \hline
${\bold{D}}^{\bold{dv}}=(a_1,{\cdots},a_{max})$ & \begin{tabular}[c]{@{}l@{}}$|\bold{dv}|$-sized vector defining the numbers of protograph VNs, where $a_i$ is the \\ number of protograph VNs of degree $l_i$  and $\bold{dv}$ is a set of VN degrees that \\exist in the protograph, i.e.,  $\bold{dv}=\{l_1,l_2,{\cdots},l_{max}\}$\end{tabular} \\ \hline
$\rho$ & Random puncturing ratio of the entire protograph \\\hline
$\rho_{d} $  & Random puncturing ratio for partially doped VNs in the protograph\\\hline
\end{tabular}%
\label{tab:1}
}
\end{table*}
In this section, we introduce some notations and concepts of a binary erasure channel, protograph LDPC codes, and the construction method of protograph doped GLDPC codes. The EXIT analysis and the decoding process of protograph doped GLDPC codes are also briefly introduced. The notations mainly used throughout the paper are summarized in Table 1. 
\subsection{Protograph LDPC Code and BEC}
Let $\bold{x}=\{x_1,{\cdots},x_k\}, x_i \in \{0,1\}$ be a $k$-bit binary message vector, which is encoded via an $(n,k)$ linear code, forming an $n$-bit codeword $\bold{c}=\{c_1,{\cdots},c_n\}, c_i \in \{0,1\}$. 
The codeword passes through a memoryless BEC, where each bit is either erased with a probability $\epsilon$ or correctly received.\\
\indent Protograph LDPC codes~\cite{Thorpe} are defined by a relatively small bipartite graph $G=(V,C,E)$ representing a protograph, where $V=\{v_1,{\cdots},v_{n_v}\}$ is a set of VNs and $C=\{c_1,{\cdots},c_{n_c}\}$ is a set of check nodes (CNs).
Let $E$ be a set of edges $e$, where $e=(v,c)$ connects a VN $v \in V$ and a CN $c \in C$. 
The bipartite graph can also be expressed in terms of an $n_c \times n_v$-sized base matrix ${\bold{B}}_{n_c \times n_v}=\{b_{i,j}\},i\in[n_c],j\in[n_v]$, where $b_{i,j} \in \{ 0,1,2,{\cdots} \}$ and $[A]$ is a set of positive integers less than or equal to a positive integer $A$. 
The rows represent the CNs and the columns represent the VNs in the protograph. 
Each entry $b_{i,j}$ of the base matrix represents the number of edges connected between a VN and a CN.
If there are no edges connected between $v_j$ and $c_i$, the entry $b_{i,j}$ is zero. 
The variable (check) node degree $deg(v_j)$ ($deg(c_i)$) is defined as the number of edges incident to itself. 
A protograph LDPC code is constructed by copy-and-permute operation of $G$. 
The bipartite graph $G$ is copied by the lifting factor $N$ and copies of each edge $e=(v,c) \in E$ are permuted among copies of $v$ and $c$. 
In general, the large value of $N$ guarantees the sparseness of the code. 

\subsection{{Construction of Protograph Doped GLDPC Codes} \rm{\cite{QC_GLDPC}}}
\begin{figure*}
    \centering
\includegraphics[draft=false,width=12cm]{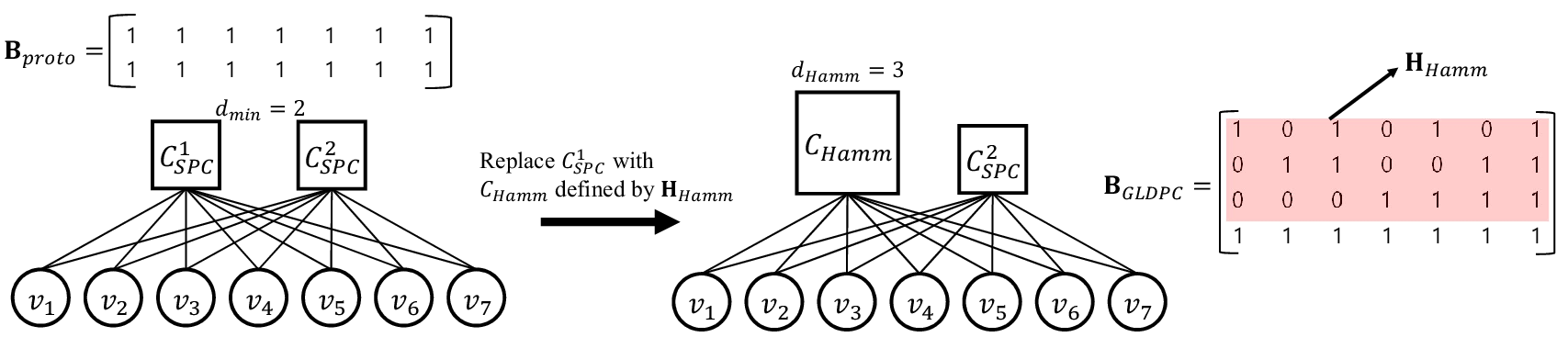}\label{fig:conven1}
    \caption{{An example of protograph doped GLDPC code construction following} \cite{QC_GLDPC} {by replacing an SPC node with a GC node using the $(7,4)$ Hamming code as the component code.}}
    \label{fig:conven_construction}
\end{figure*}
Conventionally, a protograph doped GLDPC code ensemble is constructed by replacing (doping) a CN of a protograph with a GC node that has a parity-check constraint from an $(n_i,k_i,d_{min}^i)$ linear code (component code), where $n_i$ ($k_i$) is the code length (dimension) and $d_{min}^i$ is the minimum distance of the component code for a CN $c_i$.
The condition for replacement is that the CN degree should be exactly equal to the length of the component code, i.e., $deg(c_i)=n_i$. 
Note that the original CN has the parity-check constraint of an $(n_i,k_i)=(deg(c_i),deg(c_i)-1)$ SPC code. 
The code rate $R$ of protograph doped GLDPC codes is $R=1-\frac{m_{proto}}{n_v}$, where $m_{proto}=\sum_{i=1}^{n_c}(n_i-k_i)$.
While the minimum distance of an SPC node is 2, the VNs connected to the GC node are protected by parity-check constraints of the component code with the minimum distance larger than two. Fig.~\ref{fig:conven_construction} shows the protograph doped GLDPC code of the code rate ${3}/{7}$ by replacing an SPC node with the $(7,4)$ Hamming code constraints.
\subsection{PEXIT Analysis and Decoding Process of Protograph Doped GLDPC Codes}
\begin{algorithm}[t]
 \caption{The PEXIT analysis of a protograph doped GLDPC code \cite{EXIT_Proc}}
 \label{Alg:PEXIT}
 \begin{algorithmic}[1]
\STATE \textbf{Step 1) Initialization} \\
   Initialize $I_{ch}(j)=1-\epsilon$ for $j \in [n_v]$.
\STATE \textbf{Step 2) Message update from VN to CN } \\
    Update $I_{EV}(i,j)=1-\epsilon \prod_{t\in N(v_j)}\bigl( 1-I_{AV}(t,j) \bigr)^{\delta(t,j)}$ for all $j\in[{n_v}]$, where $\delta(t,j)=b_{t,j}$ for $t\neq i $ and $\delta(t,j)=b_{t,j}-1$ for $t=i$. Further, $I_{EV}(i,j)=0$ if $b_{i,j}=0$. If $c_i$ is an SPC node, $I_{AV}(i,j)=I_{EC}(i,j)$ and if $c_i$ is a GC node, $I_{AV}(i,j)=I_{EGC}(i,j)$. 
\STATE \textbf{Step 3) Message update from CN to VN}  \\
    For all $i$, if $c_i$ is an SPC node, go to \textbf{Step 3-1)} and if $c_i$ is a GC node, go to \textbf{Step 3-2)}.  \\
    \hspace*{\algorithmicindent}  \textbf{Step 3-1)} 
     $I_{EC}(i,j)=\prod_{t\in N(c_i)}I_{AC}(i,t)^{\delta(i,t)}$, where $\delta(i,t)=b_{i,t}$ for $t\neq i $ and  \\\hspace*{\algorithmicindent} $\delta(i,t)=b_{i,t}-1$  for $t=j$. Further, $I_{AC}(i,t)=I_{EV}(i,t)$. \\
    \hspace*{\algorithmicindent} \textbf{Step 3-2)}  
     For all $j \in$ $N(c_i)$, compute 
     \begin{align}
    I_{EGC}(i,j)=&\frac{1}{n_i}\sum_{h=1}^{n_i}\bigl(1-I_{AGC}(i)\bigr)^{h-1}\bigl(I_{AGC}(i)\bigr)^{n_i-h} \nonumber\\
    &\times[h  \tilde{e}_h-(n_i-h+1) \tilde{e}_{h-1}],
    \label{eq:I_EGC}
    \end{align}
   \hspace*{\algorithmicindent}\hspace*{\algorithmicindent} where $I_{AGC}(i)=\frac{1}{n_i}\sum_{j\in N(c_i)}b_{i,j}\times I_{EV}(i,j)$.
\STATE \textbf{Step 4) APP mutual information computation}\\
    For all $j \in [n_v], I_{APP}(j)= 1-\epsilon \prod_{t\in N(v_j)}\bigl(1-I_{AV}(t,j)\bigr)^{b_{t,j}}$. If $c_t$ is an SPC node, $I_{AV}(t,j)=I_{EC}(t,j)$ and if $c_t$ is a GC node, $I_{AV}(t,j)=I_{EGC}(t,j)$. 
\STATE \textbf{Step 5) Convergence check of VNs} \\Repeat \textbf{Step 2)--4)} until $I_{APP}(j)=1,$ for all $j \in [n_v]$.
 \end{algorithmic}
 \end{algorithm}

The asymptotic performance of the protograph doped GLDPC codes is evaluated by the PEXIT analysis.  
The PEXIT analysis tracks down the mutual information of extrinsic messages and a priori error probabilities of the VNs, CNs, and GC nodes of protograph GLDPC codes. 
{For an exact PEXIT analysis, tracking down each mutual information corresponding to edges of the component code is needed, i.e., multi-dimensional EXIT computation} \cite{exact_EXIT}. 
{However, in terms of code optimization, where lots of EXIT computation is required, it is beneficial to reduce the complexity of the EXIT computation in GC nodes by averaging the a priori and extrinsic mutual information of the GC nodes.}
The EXIT and PEXIT analyses of the protograph doped GLDPC codes over the BEC in terms of average mutual information are given in \cite{EXIT_Proc, EXIT_TIT,PEXIT}.\\
\indent The PEXIT process is given in Alg.~1. Let $I_{ch}(j)$ be the channel information from the erasure channel for the protograph VN $v_j$. In addition, $I_{EV}(i,j)$  $(I_{EC}(i,j))$ is the extrinsic information sent from $v_j$ ($c_i$) to $c_i$ ($v_j$) and $I_{AV}(i,j)$ $(I_{AC}(i,j))$ is the a priori mutual information of $v_j$ ($c_i$) sent from $c_i$ ($v_j$), where $c_i$ is an SPC node. 
For GC nodes, we use the notations $I_{AGC}(i)$ and $I_{EGC}(i,j)$ for a priori and extrinsic information.
Let $N(c_i)$ ($N(v_j)$) be a set of variable (check) nodes incident to $c_i$ ($v_j$), i.e., neighborhood of $c_i$ ($v_j$). 
Finally, $I_{APP}(j)$ is a posteriori probability of $v_j$.
To explain (\ref{eq:I_EGC}) in Alg.~1, if $c_i$ is a GC node with the $(n_i,k_i)$ Hamming code, the PEXIT of the GC node is computed from a closed form using the property of the simplex code, which is the dual code of a Hamming code. 
Also, $I_{AGC}(i)=\frac{1}{n_i}\sum_{j\in N(c_i)}b_{i,j}\times I_{EV}(i,j)$ is the average a priori mutual information for a GC node to compute the PEXIT message. 
In (\ref{eq:I_EGC}), we have $$\tilde{e}_h=\sum_{t=1}^{h}t\sum_{u=0}^{t-1}(-1)^{u}2^{\binom{u}{t}}{k_i\brack t}{t\brack u}\binom{2^{t-u}}{h}.$$ 
For two positive integers $a$ and $b$, we also have $\binom{a}{b}=\prod_{i=0}^{b-1}\frac{a-i}{b-i}$ and ${a \brack b}=\prod_{i=0}^{b-1}\frac{2^a-2^i}{2^b-2^i}$, where $\binom{a}{0}=1$ and ${a\brack 0}=1$.
The PEXIT process searches for the minimum $\epsilon$ to successfully decode, i.e., $I_{APP}(j)=1,$ for all $j\in[n_v]$, in an asymptotic sense. \\
\indent Now, we briefly explain the decoding  process  of GLDPC  codes over the BEC~\cite{SC-LDPC3}.
The VNs process the conventional message-passing decoding over the BEC by sending correct extrinsic messages to the CNs if any of the incoming bits from their neighborhood is not erased. 
The SPC nodes send correct extrinsic messages to the VNs if all of their incoming messages are correctly received, and send erasure messages otherwise. 
In this paper, the decoding of GC nodes is processed by the maximum likelihood (ML) decoder. 
For each iteration, a GC node $c_i$ with the $(n_i,k_i)$ component code receives the set of erasure locations $\{ e_i \}$ from $N(c_i)$. 
Let $H_{GC}$ be the parity-check matrix (PCM) of the component code and $H_{e}$ be the submatrix of $H_{GC}$ indexed with $\{ e_i \}$. 
The decoder computes the Gaussian-elimination operation of $H_e$, making it into a reduced row echelon form $H_e^{reduced}$. 
If rank($H_{e}^{reduced}$)$ = |\{e_i\}|$, the GC node solves all the input erasures and otherwise, the decoder corrects the erasures corresponding to the rows with weight 1 from $H_e^{reduced}$. 
The decoding complexity can be further reduced if the GC node exploits bounded distance decoding; however, the degradation of asymptotic performance is not negligible, as shown in~\cite{IT_GLDPC}.

\vspace{10pt}
\section{The Proposed Protograph-based PD-GLDPC Code}
\begin{figure}
    \centering
    \subfloat[Conventional protograph GLDPC code]{
\includegraphics[draft=false,width=7cm]{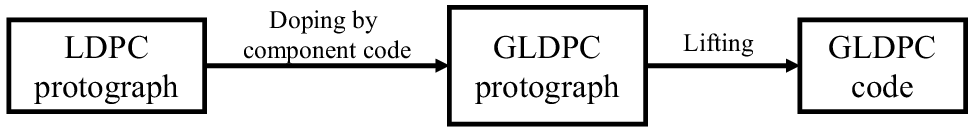}\label{fig:blockdiagram1}
}
 \subfloat[Conventional random ensemble GLDPC code]{
\includegraphics[draft=false,width=7cm]{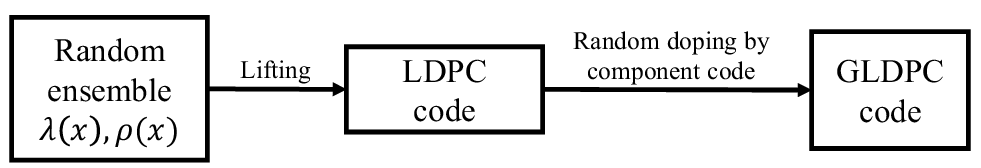}\label{fig:blockdiagram_random}
}
\\
\subfloat[Proposed protograph-based PD-GLDPC code]{
\includegraphics[draft=false,width=7cm]{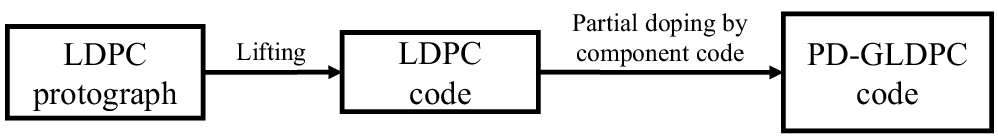}\label{fig:blockdiagram2}
}
    \caption{Block diagram of the construction process of conventional GLDPC codes and the proposed protograph-based PD-GLDPC codes.}
    \label{fig:blockdiagram}
\end{figure}
In this section, a new construction method of protograph-based GLDPC codes is proposed. 
While the conventional protograph GLDPC codes are constructed by replacing some protograph SPC nodes in the original protograph by GC nodes using the component code, the proposed protograph-based PD-GLDPC code is constructed by adding the GC nodes for the subset of variable nodes using component codes after the lifting process of the original protograph, where each GC node is connected to the variable nodes copied from single protograph variable node. 
A block diagram of the construction process of both codes together with the conventional random GLDPC code is given in Fig.~\ref{fig:blockdiagram}.

\subsection{Construction Method of Protograph-based PD-GLDPC Code}

\begin{figure}
    \centering
    \includegraphics[draft=false,width=13cm]{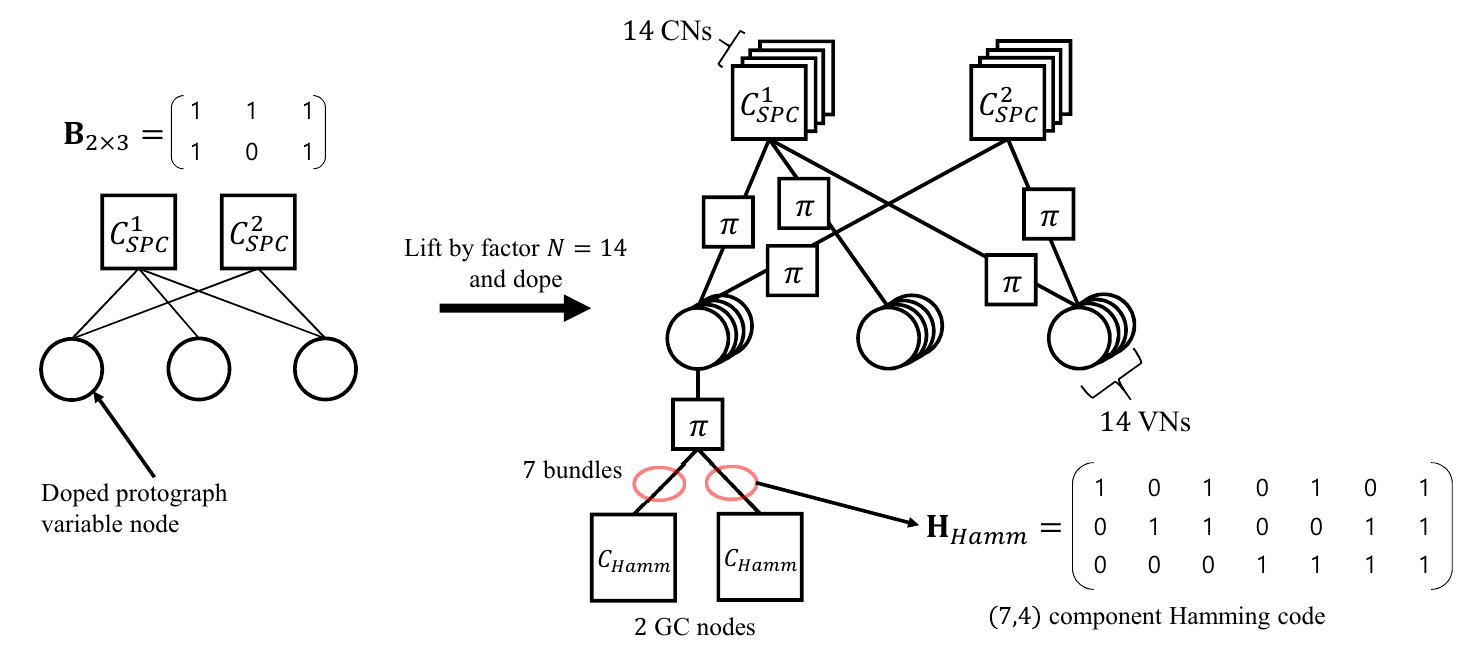}
    
    \caption{An example of a proposed $({\bold{B}}_{2 \times 3},7,4,14,1)$ protograph-based PD-GLDPC code construction, where ${\bold{B}}_{2 \times 3}=[1 \, 1\, 1;1\, 0\, 1]$ and $\pi$ is the $14\times 14$ sized permutation matrix.}
    \label{fig:proposed_ex}
\end{figure}
First, we define the partial doping for variable nodes using the addition of GC nodes to a lifted protograph, that is, the addition of rows for the parity check matrix by the component code, where each GC node is incident to variable nodes copied from single protograph variable node. 
Also, we define a partially doped protograph variable node as a protograph variable node incident to the added GC nodes. 
While the term doping in conventional GLDPC codes is used in the perspective of check nodes, we use the term partial doping in the perspective of variable nodes.
Let ${\bold{B}}_{n_c \times n_v}$ be an $n_c \times n_v$ base matrix, where some protograph variable nodes are partially doped with a $(\mu,\kappa)$ component code after the lifting process. 
Let $x=0,1,2,{\cdots}$ be the  number of the partially doped protograph variable nodes, where each protograph variable node is doped by $N / \mu$ component codes after the lifting process. 
Then, a proposed protograph-based PD-GLDPC code is defined with parameters $({\bold{B}}_{n_c \times n_v},\mu,\kappa,N,x)$. 
We assume that the component code used in the paper is the $(\mu,\kappa)$ Hamming code and that $\mu$ divides the lifting factor $N$ such that $N=\mu \beta$, where $\beta$ is a non-negative integer.
The variable nodes copied from $x$ protograph variable nodes in the base matrix are partially doped by the GC nodes.
That is, in the proposed PD-GLDPC code construction, the $N/\mu$ GC nodes are connected to the $N$ variable nodes lifted from each protograph variable node.
Thus, the proposed construction method can choose the protograph variable nodes to protect by partial doping. 
An example of the proposed construction is given in Fig.~\ref{fig:proposed_ex}, which illustrates the doping process by a $(7,4)$ component Hamming code over a $2\times3$ base matrix. 
\begin{figure}
    \centering
    \subfloat[PCM of $\beta$ GC nodes doped for single protograph variable node ]{\includegraphics[draft=false,width=9cm]{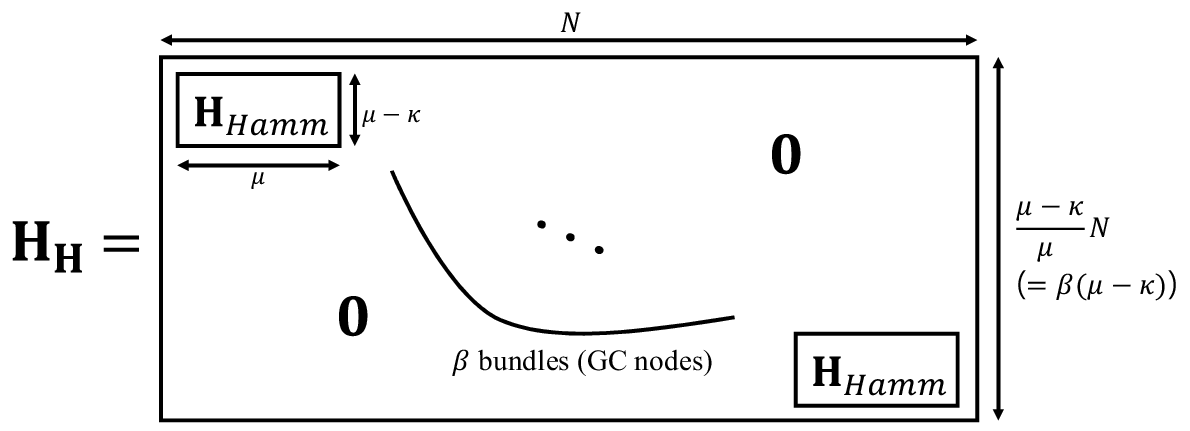}}
    
    \subfloat[PCM of $x$ partially doped PD-GLDPC code ]{\includegraphics[draft=false,width=13cm]{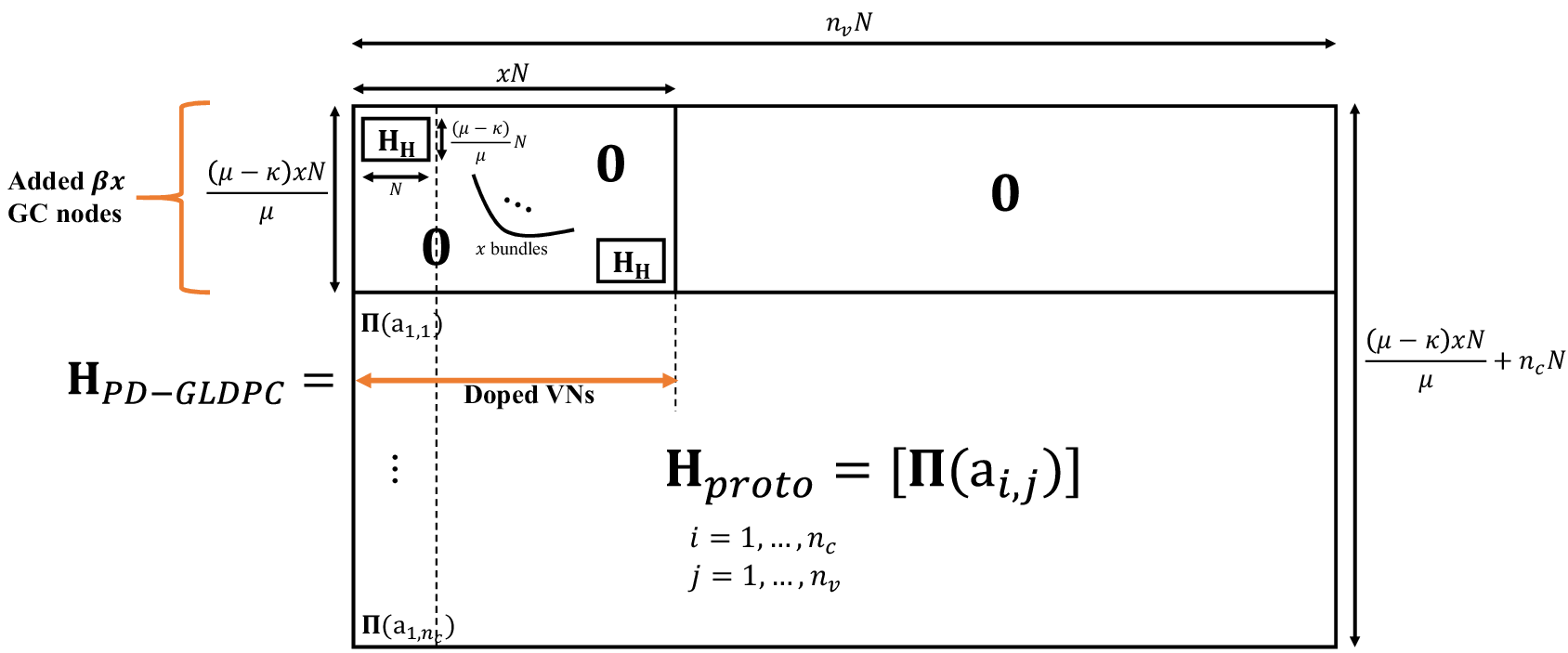}}
    \caption{PCM of a protograph-based PD-GLDPC code.
    }
    \label{fig:PCM_GLDPC}
\end{figure}
\indent The basic concept of the proposed construction is to focus on the protection of VNs lifted from single protograph VN. {Although only a portion of VNs from a single protograph VN can be partially doped, for the exactness of EXIT computation and the typical minimum distance analysis, we have limited the doping process over the entire VNs lifted from a single protograph VN.}  
Since $N$ is the multiple of the component code length, all the VNs lifted from single protograph VN can be protected by using $\beta$ GC nodes. 
{A simple example for the PCM for $\beta$ GC nodes, connected to the VNs lifted from single protograph VN, $\textbf{H}_{\textbf{H}}$ is shown in Fig.}~\ref{fig:PCM_GLDPC}(a), {where $\textbf{H}_{Hamm}$ is the PCM of the Hamming code. 
Although the PCM of the Hamming code can be applied randomly, a trivial representation of applying generalized constraints sequentially is given.
The constructed PD-GLDPC code has a PCM ${\bold{H}}_{PD{\text{-}}GLDPC}$ as in Fig.}~\ref{fig:PCM_GLDPC}(b), {where the upper part is the PCM of the added $\beta x$ GC nodes and the lower part ${\bold{H}_{proto}}$ refers to the PCM of the LDPC code lifted from the original protograph. }
Intuitively, $\textbf{H}_{\textbf{H}}$ represents the PCM for each partially doped VN in the protograph and thus, the $x=|\mathcal{X}|$ bundles of matrices are diagonally appended to the PCM of the PD-GLDPC code.
Since the doping proceeds after the lifting process, PD-GLDPC codes cannot be expressed in terms of a protograph. 
{{We define the doping ratio $\nu$ as the portion of GC nodes over the entire constraint nodes, i.e., $\nu=\frac{x\beta }{x\beta+n_c N}.$}}
{Also, we define the doping granularity as the minimum number of protograph VNs needed for doping. For the protograph doped GLDPC codes with $(\mu,\kappa)$ component code, the doping granularity is $\mu$, whereas the proposed PD-GLDPC code has doping granularity 1. The finer doping granularity of the PD-GLDPC codes allows the construction of protograph-based GLDPC codes with the higher rate.}\\
\subsection{The PEXIT Analysis of Protograph-based PD-GLDPC Codes}
\begin{figure}
    \centering
    \includegraphics[draft=false,width=7cm]{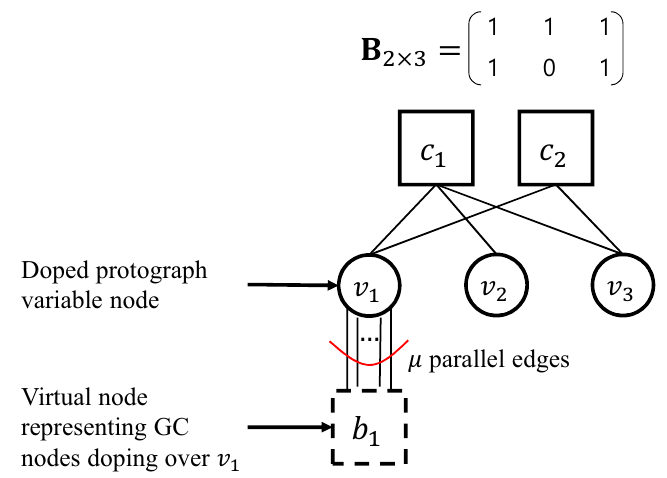}
    \caption{An example of the Tanner graph representation of the proposed PD-GLDPC code from the protograph ${\bold{B}}_{2 \times 3}=[1 \, 1 \, 1;1 \, 0 \, 1]$, where $x=1$.
    }
    \label{fig:Virtual}
\end{figure}
\indent The PEXIT of the proposed protograph-based PD-GLDPC codes is similar to that of the conventional protograph GLDPC codes in Alg.~1 except for the EXIT of a GC node. 
Since the incoming mutual information of each GC node is obtained from only one protograph variable node in the proposed code, the average mutual information sent to each GC node is the same as the extrinsic message of the protograph variable node connected to the GC node. 
Let $b_j, j\in [x]$ be the virtual node representing the set of $\beta$ GC nodes connected to the protograph variable node $v_j$. An example of the representation of a virtual node over a protograph is given in Fig.~\ref{fig:Virtual}.
Note that although $b_j$ is not a protograph node itself, it is possible to compute the PEXIT of a protograph-based PD-GLDPC code.
Also, let $I_{EV}^{(b_j)}(j)$ be the extrinsic information from $v_j$ to $b_j$  expressed as 
$$I_{EV}^{(b_j)}(j)=1-\epsilon \prod_{t\in N(v_j)}(1-I_{AV}(t,j)),j\in[x].$$  
Since $b_j$ is solely connected to $v_j$, the index term for the extrinsic information from $v_j$ to $b_j$ is expressed by the notation of $j$ only. In order to compute the EXIT of $b_j$, let $I_{AGC}^{(b_j)}(j)$ and $I_{EGC}^{(b_j)}(j)$ be the a priori and extrinsic mutual informations of $b_j$, respectively. Note that the EXIT of each GC node is computed from single a priori mutual information to process single value of extrinsic mutual information for the neighboring variable nodes. For the conventional protograph GLDPC codes, the a priori mutual information of a protograph GC node is computed by averaging the extrinsic mutual information of its neighboring variable nodes. However, for the proposed protograph-based PD-GLDPC codes, since $b_j$ receives the extrinsic mutual information of $v_j$ only, it is clear that $I_{AGC}^{(b_j)}(j)=I_{EV}^{(b_j)}(j),\,j\in[x]$. 
We also compute the extrinsic mutual information from $b_j$ to $v_j$ denoted as $I_{EGC}^{(b_j)}(j)$ using (\ref{eq:I_EGC}), given the a priori mutual information $I_{AGC}^{(b_j)}(j)$, which is given as
\begin{equation}
  I_{EGC}^{(b_j)}(j)=\frac{1}{\mu}\sum_{h=1}^{\mu}\bigl(1-I_{AGC}^{(b_j)}(j)\bigr)^{h-1}\bigl(I_{AGC}^{(b_j)}(j)\bigr)^{\mu-h}[h  \tilde{e}_h-(\mu-h+1) \tilde{e}_{h-1}].  
\end{equation}

Note that for the proposed protograph-based PD-GLDPC codes, the a priori (extrinsic) EXIT of the GC node is computed from the extrinsic (a priori) EXIT of single protograph variable node. While the EXIT of variable nodes and SPC nodes for the proposed protograph-based PD-GLDPC codes is the same as that of the conventional protograph GLDPC codes described in Alg.~1, the EXIT of the GC nodes in the proposed codes is changed to (2) whereas the conventional protograph GLDPC codes use (1). \\
\indent In general, as the portion of degree-2 variable nodes in the LDPC codes increases, the asymptotic performance is enhanced~\cite{Double_protograph}, but their minimum distance decreases and then the error floor becomes worse.
Although the proposed protograph-based PD-GLDPC code construction method enables the partial doping for any variable nodes in the given protographs, we focus only on partial doping for degree-2 variable nodes as follows.
First, we construct the original base matrix ${\bold{H}_{proto}}$ with the large portion of degree-2 variable nodes.
Then, we partially dope some of the protograph variable nodes of degree-2 to increase the minimum distance and improve their performance. 
Thus, regular protographs of degree-2 variable nodes and irregular protographs with many degree-2 variable nodes are used for the construction of the proposed protograph-based PD-GLDPC codes. 
In terms of irregular ensemble LDPC codes, a large portion of degree-2 variable nodes enables the LDPC code to achieve the capacity approaching performance~\cite{deg2_thres_IT}. 
On the other hand, by reasonably selecting the number of partially doped variable nodes for degree-2, the property of the linear minimum distance growth with the length of the LDPC code can be guaranteed. 
Thus, when we design the proposed protograph-based PD-GLDPC codes, balancing the partial doping over degree-2 variable nodes enables both the existence of a typical minimum distance and a good asymptotic performance. 

\subsection{Condition for Typical Minimum Distance of Protograph-based PD-GLDPC Code}
The existence of a typical minimum distance in the given LDPC codes guarantees that their minimum distance grows linearly with the block length of the code in an asymptotic sense \cite{MCT}. 
It was proved in \cite{Tdmin_protograph} that a protograph LDPC code has a typical minimum distance if there is no cycle consisting of only degree-2 variable nodes in the protograph. 
Furthermore, in \cite{Tdmin_GLDPC}, the condition for a typical minimum distance of the conventional protograph GLDPC codes was given as follows. 
Assuming that there are no degree-2 variable nodes with double edges, i.e., no type 1 degree-2 variable nodes defined in \cite{Tdmin_GLDPC}, the neighborhoods of a check node $c_i$ that satisfy $d^{i}_{min} \geq N^{(dv2)}(c_i)$ are removed along with its edges, where $N^{(dv2)}(c_i)$ is the number of degree-2 variable nodes among the neighborhoods of $c_i$. 
This process is repeated until no further degree-2 variable nodes remain.
Then, the GLDPC code has a typical minimum distance if all degree-2 variable nodes are removed. \\
\indent The proposed protograph-based PD-GLDPC code also has a similar approach to that of the conventional protograph GLDPC codes in \cite{Tdmin_GLDPC}.
However, since a GC node of the proposed protograph-based PD-GLDPC code is not well defined by a protograph node, the derivation of the weight enumerator of the proposed codeword is quite different from that of the conventional GLDPC code.
Thus, the condition for the existence of the typical minimum distance of the proposed protograph-based PD-GLDPC codes is slightly different from that of the conventional protograph LDPC code. 
That is, the degree-2 variable nodes to be partially doped are initially excluded before the degree-2 variable node removing process stated in \cite{Tdmin_protograph}, because those variable nodes should be changed to the variable nodes of degrees higher than two by partial doping. Thus, we can regard the degree-2 variable nodes to be partially doped as the variable nodes with higher degrees.
Then, we have the following theorem for the proposed protograph-based PD-GLPDC codes.

\begin{theorem}
For a $({\bold{B}}_{n_c \times n_v},\mu,\kappa,N,x)$ protograph-based PD-GLDPC code, if the undoped degree-2 variable nodes in the protograph have no cycles among themselves, the proposed protograph-based PD-GLDPC code has a typical minimum distance.
\proof The proof is given in Appendix A.
\end{theorem}
The existence of the typical minimum distance of the proposed protograph-based PD-GLDPC code guarantees that the minimum distance of the proposed code grows linearly with the code length of the LDPC code, and thus the proposed code has low error floor for the large code length. In the next section, we use Theorem 1 as the constraint to optimize the protograph in order to guarantee the existence of the typical minimum distance of the proposed protograph-based PD-GLDPC code. 

\subsection{Comparison between the Proposed PD-GLDPC Code and the Conventional Protograph GLDPC Code}
The main difference between the proposed  protograph-based PD-GLDPC codes and the conventional protograph GLDPC codes is the perspective of doping. 
While the conventional protograph GLDPC code replaces an entire row, i.e., a protograph check node by the parity check matrix of the component code, the proposed protograph-based PD-GLDPC code appends some rows incident to the variable nodes copied from single protograph variable node. 
The focus of the conventional protograph GLDPC code is to choose a certain protograph check node, whereas the protograph-based PD-GLDPC code is focusing on choosing which protograph variable nodes are further protected by partial doping.
The constraint for the conventional protograph GLDPC code is that the check nodes to be replaced should have the degree equal to the component code length, while the constraint for the proposed protograph-based PD-GLDPC codes is that the lifting size of a protograph should be the multiple of the component code length. \\
\indent The selection of the check nodes to be replaced in the conventional protograph GLDPC codes is generally difficult since it requires a combinatory search of either the best threshold, minimum distance, or convergence speed.
Also, since many SPC nodes of a given base matrix connect the variable nodes with various degrees, it is very difficult to select the variable nodes to be partially doped for further protection in the conventional protograph GLDPC codes. \\
\indent Furthermore, the conventional protograph GLDPC code construction has two limitations in terms of PEXIT analysis and doping granularity.
First, the PEXIT analysis of the conventional protograph GLDPC codes is not accurate because the PEXIT of the GC node is derived from averaged EXIT of $\mu$ (component code length) protograph variable nodes. 
Using a MAP-oriented computation as in Alg. 1, the GC node outputs a single value that represents the extrinsic mutual information for the $\mu$ protograph variable nodes. 
Thus, a GC node averages up the a priori mutual information of the $\mu$ neighborhood protograph variable nodes and outputs uniform extrinsic mutual information, which is similar to the EXIT analysis of a random ensemble LDPC code \cite{EXIT_TIT}. 
Since each GC node receives a priori inputs from $\mu$ different protograph variable nodes, the EXIT analysis is not accurate. 
Thus, the higher variance of the a priori mutual information from the average, the greater the deviation of the code between the threshold and the actual decoding performance. 
On the other hand, since the EXIT of a GC node in the proposed protograph-based PD-GLDPC code requires the a priori mutual information of the same protograph variable node, there is no deviation since the $\mu$ mutual information has the same value, which makes the PEXIT of the proposed code more accurate.\\
\indent The second limitation is that the conventional protograph GLDPC codes have large doping granularity of protograph variable nodes compared to that of the proposed one.
By replacing a single protograph GC node by a component code with parameters $(\mu,\kappa)$, the $\mu$ protograph variable nodes are doped. Whereas, for every partial doping of $\beta$ GC nodes in the proposed protograph-based PD-GLDPC code, the variable nodes copied from single protograph variable node are partially doped. In other words, the doping granularity is one, which is smaller than the conventional protograph GLDPC codes. \\
\indent In summary, compared to the conventional protograph GLDPC codes, the proposed protograph-based PD-GLDPC code is more accurate in the PEXIT analysis and has the smaller doping granularity.

\vspace{10pt}
\section{Optimization of PD-GLDPC Codes}
{In this section, we introduce two optimization methods for the PD-GLDPC codes.} The first subsection illustrates the construction method of protographs from the degree distribution of a random LDPC code ensemble in order to conduct comparison between LDPC codes and PD-GLDPC codes under the same degree distribution. The second subsection shows the optimization method of the protograph using the differential evolution algorithm in order to conduct comparison between LDPC codes and PD-GLDPC codes without any constraints. 

\subsection{Differential Evolution-Based Code Construction from the Degree Distribution of Random LDPC Code Ensembles}
\indent In general, as the portion of degree-2 VNs in the LDPC codes increases, the asymptotic performance is enhanced~\cite{Double_protograph}, but their minimum distance decreases and then the error floor becomes worse.
For the construction of PD-GLDPC codes in this subsection, we exploit the balance of the portion of degree-2 VNs, where we focus on the partial doping only for degree-2 VNs. The brief construction method is as follows.
First, we construct the original base matrix ${\bold{B}_{n_{c}\times n_v}}$ with the large portion of degree-2 VNs.
Then, we partially dope some of the protograph VNs of degree-2 to increase the minimum distance and improve their performance. 
Thus, irregular protographs with several degree-2 VNs are used for the construction of the proposed PD-GLDPC codes. 
In terms of irregular LDPC code ensembles, a large portion of degree-2 VNs enables the LDPC code to achieve the capacity approaching performance~\cite{deg2_thres_IT}. 
On the other hand, by reasonably selecting the number of partially doped VNs of degree-2, the property of the linear minimum distance growth with the length of the LDPC code can be guaranteed. 
Thus, when we design the proposed PD-GLDPC codes, balancing the partial doping over degree-2 VNs enables both the existence of a typical minimum distance and a good asymptotic performance. 
Optimization of irregular protograph LDPC code ensembles is made by initially obtaining the degree distribution of the random LDPC code ensemble using differential evolution \cite{Differential evolution} and constructing the protograph via the progressive edge growth (PEG) \cite{PEG} algorithm for the construction of the proposed PD-GLPDC codes from
irregular protographs. 
In this subsection, in order to make the CN degrees as even as possible, we try to construct the protograph from the degree distribution of a random LDPC code ensemble. 
We define $G_c$ as the optimized protograph of the conventional LDPC code and $G_p$ as the initial irregular protograph that is used to construct the PD-GLDPC code. 
That is, we can regard $G_p$ as the protograph corresponding to $H_{proto}$ in Fig.~\ref{fig:PCM_GLDPC}. 
In order to compare FER performances of the conventional LDPC code and the proposed PD-GLDPC code under the same degree distribution, $G_c$ is constructed to have the same VN degree distribution as the PD-GLDPC code constructed from $G_p$ after lifting by $N$.\\
\indent Let $\lambda_{G_c}(x)$ and $\rho_{G_c}(x)$ be the VN and CN degree distributions of an irregular LDPC code ensemble to construct $G_c$, which is the optimized protograph for the conventional LDPC codes. 
In this subsection, we assume the degree distributions $\lambda_{G_c}(x)=\lambda_{2}x+\lambda_{3}x^2+\lambda_{4}x^3+\lambda_{5}x^4+\lambda_{6}x^5+\lambda_{l}x^{l-1}$ and $\rho_{G_c}(x)=\rho_{r-1}x^{r-2}+\rho_{r}x^{r-1}$, where $\lambda_{i}$ and $\rho_i$ are the portions of edges of VNs and CNs of degree-$i$. 
Using the optimized degree distributions of $\lambda_{G_c}(x)$ and $\rho_{G_c}(x)$, a protograph $G_c$ is constructed by the PEG algorithm.
For the description of the protographs that construct the conventional LDPC codes and the proposed PD-GLDPC codes, let ${\bold{D}}^{\bold{dv}}=(a_1,{\cdots},a_{max})$ be a $|\bold{dv}|$-sized vector defining the numbers of protograph VNs, where $a_i$ is the number of protograph VNs of degree $l_i$ and $\bold{dv}=\{l_1,l_2,{\cdots},l_{max}\}$ is a set of VN degrees that exist in the protograph. \\
\begin{algorithm}[t]
 \caption{Construction of $G_c$ and the PD-GLDPC code}
 \label{Alg:irr}
 \begin{algorithmic}[1]
 \renewcommand{\algorithmicrequire}{\textbf{Input:} }
 \REQUIRE $\mu$, $\kappa$, $n_v$, $n_c$, $R$, $l$, $r$, $y_{max}$
 \renewcommand{\algorithmicrequire}{\textbf{Output:} }
 \REQUIRE $y^{opt}$, $G_c$, $G_p$
 \STATE \textbf{Step 1) Optimize degree distribution of $G_c$}\\ Optimize $\lambda_{G_c}(x)=\lambda_{2}x+\lambda_{3}x^2+\lambda_{4}x^3+\lambda_{5}x^4+\lambda_{6}x^5+\lambda_{l}x^{l-1}$ and $\rho_{G_c}(x)=\rho_{r-1}x^{r-2}+\rho_{r}x^{r-1}$ using differential evolution under constraints (a)$\sim$(c): 
 \begin{enumerate}
     \item[(a)] rate constraint $R=1-\frac{\int_{0}^{1}\rho_{G_c}(x)dx}{\int_{0}^{1}\lambda_{G_c}(x)dx},0\leq \lambda_{i} \leq 1, 0 \leq \rho_{i} \leq 1$
     \item[(b)] typical minimum distance constraint $\frac{\lambda_{2}/2}{\Sigma} \times n_v \leq n_c-1-y_{max}(\mu-\kappa) \leftrightarrow \lambda_{2} \leq \frac{2\Sigma \{n_c-1-y_{max}(\mu-\kappa) \}}{n_v}$
     \item[(c)] $G_{p}$ existence constraint $\lambda_{3} \geq \frac{12\Sigma y_{max}}{n_v},\lambda_{4} \geq \frac{24\Sigma y_{max}}{n_v},\lambda_{5} \geq \frac{20\Sigma y_{max}}{n_v},\lambda_{6} \geq \frac{6\Sigma y_{max}}{n_v}$
 \end{enumerate}
\STATE \textbf{Step 2) Construction of $G_c$} \\
    From the optimized degree distribution and the random PEG algorithm, construct $G_c$ defined as $\bold{D}^{(2,3,4,5,6,l)}=(a,b,c,d,e,f)$ guaranteeing a typical minimum distance.
\STATE \textbf{Step 3) Optimization of $G_p$} \\
     For each $y=1,2,{\cdots},y_{max}$, construct $G_p$ defined as $\bold{D}^{(2,3,4,5,6,l)}=(a+15y,b-4y,c-6y,d-4y,e-y,f)$ and choose $y^{opt} \in \{y\}$ with the best threshold.  
\STATE \textbf{Step 4) Typical minimum distance check of the PD-GLDPC code} \\
     For the chosen $y^{opt}$ and $G_p$, if there exists any cycle for the submatrix induced by undoped VNs of degree-2, go to \textbf{Step 2)}. Otherwise, output $y^{opt}$ and $G_p$.
 \end{algorithmic}
 \end{algorithm}
\indent In order to make the same VN degree distributions of the LDPC codes constructed from $G_c$ and the PD-GLDPC codes constructed from $G_p$ after lifting by $N$, optimization of $\lambda_{G_c}(x)$ and $\rho_{G_c}(x)$ should be constrained by $y_{max}$, which is the maximum number of bulks of protograph VNs allowed to be partially doped in $G_p$. 
Although doping granularity for the proposed PD-GLDPC code is 1, we consider doping for bulks of protograph VNs in order to easily match the code rate and degree distribution because the purpose of this subsection is comparing FER performances between the conventional LDPC code and the proposed PD-GLDPC code under the same degree distribution.
A PD-GLDPC code is constructed by partially doping $\mu y$ protograph VNs in $G_p$. 
Construction of a PD-GLDPC code from $G_p$ is optimized by ranging the doping bulk $y$, $1\leq y \leq y_{max}$. 
That is, we search for the optimal value $y$ which maximizes the coding gain between the PD-GLDPC codes constructed from $G_p$ and the conventional protograph LDPC codes constructed from $G_c$.\\
\indent Conditions for the degree distributions in order to construct $G_c$ are derived as follows.
The conditions need to guarantee two criteria: i) the VN degree distributions of the protograph LDPC code constructed from $G_c$ and the PD-GLDPC code constructed from $G_p$ after lifting by $N$ are the same and ii) a typical minimum distance exists for both code ensembles. 
In this subsection, we assume that partial doping is conducted for the first $\mu y$ degree-2 protograph VNs without loss of generality due to randomness of the PEG algorithm. 
For the $y$ bulks of partially doped protograph VNs using the PCM of the $(15,11)$ Hamming code, the numbers of protograph VNs in $G_p$ should be $$\bold{D}^{(2,3,4,5,6,l)}=(a+15y,b-4y,c-6y,d-4y,e-y,f).$$ Given that $G_c$ is represented as $\bold{D}^{(2,3,4,5,6,l)}=(a,b,c,d,e,f)$, for the existence constraint, each element of $\bold{D}^{(2,3,4,5,6,l)}$ should be non-negative. 
The parameters $a$$\sim$$f$ are approximated by the PEG construction as $$
a \approx \floor*{n_v\frac{\lambda_{2}/2}{\Sigma}}, b \approx \floor*{n_v\frac{\lambda_{3}/3}{\Sigma}}, c \approx \floor*{n_v\frac{\lambda_{4}/4}{\Sigma}},$$$$d \approx \floor*{n_v\frac{\lambda_{5}/5}{\Sigma}}, e \approx \floor*{n_v\frac{\lambda_{6}/6}{\Sigma}},\, {\rm{and}} \, f \approx \floor*{n_v\frac{\lambda_{l}/l}{\Sigma}},$$ where $\Sigma=\int_{0}^{1}\lambda_{G_c}(x)dx$. 
For the realization of the protograph from the degree distribution using the PEG algorithm, if the summation $a+b+c+d+e+f$ is lower than $n_v$, the values of $a$$\sim$$f$ are added by 1 in order starting from the lowest VN degree until the summation is equal to $n_v$.\\
\indent If $G_c$ is determined for a given $y_{max}$ as $\bold{D}^{(2,3,4,5,6,l)}=(a,b,c,d,e,f)$, where $a+b+c+d+e+f=n_v$, $G_p$ defined by $\bold{D}^{(2,3,4,5,6,l)}=(a+15y,b-4y,c-6y,d-4y,e-y,f)$ can be constructed for $y=1,{\cdots},y_{max}$. 
By allowing the PEG algorithm of the VN degree distribution over a base matrix with size $\{n_c-(\mu-\kappa)y \} \times n_v$, both the code rate and the VN degree distributions for the LDPC codes constructed from $G_c$ and the proposed PD-GLDPC codes constructed from $G_p$ after lifting by $N$ are matched. 
We search for the value of $y$, which has the best PEXIT threshold while having a typical minimum distance. 
The optimized doping value is denoted as $y^{opt}$. The construction of $G_c$ and the PD-GLDPC code is described in Alg.~\ref{Alg:irr}.

\begin{table*}[t]
\caption{{Simulation results for optimized PD-GLDPC codes from irregular protographs using Alg. 2, where $l=20, n_v=400, R=1/2$.}}
\resizebox{\textwidth}{!}{%
\begin{tabular}{|c|c|c|c|c|}
\hline
$y_{max}$ & $\lambda_{G_c}(x),\rho_{G_c}(x)$ (threshold)                                                                                                                                  & \begin{tabular}[c]{@{}c@{}}$G_c$ protograph\\ $\bold{D}^{(2,3,4,5,6,20)}$\\ / $G_c$ threshold\end{tabular} & \begin{tabular}[c]{@{}c@{}}$G_p$ protograph\\ $\bold{D}^{(2,3,4,5,6,20)}$, $y^{opt}$\\ / PD-GLDPC threshold\end{tabular} & Coding gain \\ \hline
5         & \begin{tabular}[c]{@{}c@{}}$\lambda_{G_c}(x)=0.2049x+0.2489x^2+0.1150x^3$\\ $+0.074x^4+0.0210x^5+0.3363x^{19}$\\ $\rho_{G_c}(x)=0.9735x^7+0.0265x^8$ ($0.4815$)\end{tabular}  & \begin{tabular}[c]{@{}c@{}}$(165,134,47,23,5,26)$\\ / $0.4620$\end{tabular}                              & \begin{tabular}[c]{@{}c@{}}$(240,114,17,3,0,26)$, $y^{opt}=5$\\ / $0.4699$\end{tabular}                                 & $0.0079$    \\ \hline
10        & \begin{tabular}[c]{@{}c@{}}$\lambda_{G_c}(x)=0.1894x+0.2255x^2+0.1431x^3$\\ $+0.1191x^4+0.0357x^5+0.2872x^{19}$\\ $\rho_{G_c}(x)=0.9908x^7+0.0012x^8$ ($0.4696$)\end{tabular} & \begin{tabular}[c]{@{}c@{}}$(152,121,57,38,9,23)$\\ / $0.4523$\end{tabular}                              & \begin{tabular}[c]{@{}c@{}}$(287,85,3,2,0,23)$, $y^{opt}=9$\\ / $0.4638$\end{tabular}                                   & $0.0115$    \\ \hline
15        & \begin{tabular}[c]{@{}c@{}}$\lambda_{G_c}(x)=0.1632x+0.1758x^2+0.2143x^3$\\ $+0.1827x^4+0.0543x^5+0.2098x^{19}$\\ $\rho_{G_c}(x)=0.9940x^7+0.0060x^8$ ($0.4476$)\end{tabular} & \begin{tabular}[c]{@{}c@{}}$(131,94,86,59,14,16)$\\ / $0.4352$\end{tabular}                              & \begin{tabular}[c]{@{}c@{}}$(341,38,2,3,0,16)$, $y^{opt}=14$\\ / $0.4534$\end{tabular}                                  & $0.0182$    \\ \hline
\end{tabular}%
}
\label{tab:IRR-half}
\end{table*}
\indent The protograph of the conventional protograph LDPC code, $G_c$ is made for $y_{max}=5,10,15$ for the half-rate protograph LDPC code ensemble. 
The numerical results are summarized in Table~\ref{tab:IRR-half}, where the coding gain given for the proposed PD-GLDPC code is compared to the conventional protograph LDPC code with the equal degree distribution.

\subsection{{Optimization of PD-GLDPC Codes Using Protograph Differential Evolution}}
\begin{algorithm}[t]
 \caption{Differential evolution algorithm to design the base matrix of the PD-GLDPC codes}
 \label{Alg:proto}
 \begin{algorithmic}[1]
 \renewcommand{\algorithmicrequire}{\textbf{Input:} }
 \REQUIRE $\mu$, $\kappa$, $n_c$, $n_v$, $\mathcal{X}$, $g$, $t$, $N_p$, $p_c$, $F$, $\alpha$
 \renewcommand{\algorithmicrequire}{\textbf{Output:} }
 \REQUIRE ${\bold{B}}_{n_c \times n_v}$
 \STATE \textbf{Initialization:} Set the initial base matrices $(\bold{B}_1,{\ldots},\bold{B}_{N_p})$ each with size $n_c \times n_v$ randomly, where each entry is chosen from $\{0,{\ldots},t \}$.
 \FOR {$m=1:g$}
 \STATE \textbf{Mutation:} For each $k\in \{ 1,{\ldots},N_p \}$, the mutation matrices are created through the interpolation as follow:  
 $$[\bold{M}_k]_{i,j}= [\bold{B}_{r_1}]_{i,j}+(F+\alpha (1-F))([\bold{B}_{r_2}]_{i,j}-[\bold{B}_{r_3}]_{i,j}),$$
 where $[\bold{A}]_{i,j}$ is the $(i,j)$ element of the matrix $\bold{A}$ and indices $r_i\in [N_p],i=1,2,3$ are distinct and randomly selected. Each entry of $\bold{M}_k$ is replaced with the nearest integer in $\{ 0,{\ldots},t\}$.
\STATE \textbf{Crossover:} For each $k\in \{ 1,{\ldots},N_p \}$, create the trial matrices $\bold{M'}_k$ such that $[\bold{M'}_{k}]_{i,j}=[\bold{M}_{k}]_{i,j}$ with a probability $p_c$ and $[\bold{M'}_{k}]_{i,j}=[\bold{B}_{k}]_{i,j}$ with probability $1-p_c$. If $\bold{M'}_{k}$ contains any cycles only consisting of undoped degree-2 protograph VNs, $\bold{M'}_{k}$ is regenerated. 

\STATE \textbf{Selection:} Each base matrix in the candidates for $(m+1)$th generation is chosen between $\bold{B}_{k}$ and $\bold{M'}_{k}$. If the threshold of $\bold{B}_{k}$ is larger than $\bold{M'}_{k}$, no update is made. Otherwise, update $\bold{B}_{k}$ to $\bold{M'}_{k}$.
\ENDFOR
\STATE From $\bold{B}_k, k\in [N_p]$, choose the matrix with the best threshold value and output $\bold{B}_{n_c \times n_v}$.
 \end{algorithmic}
 \end{algorithm}

\indent In this subsection, we propose the optimization method using the differential evolution algorithm.
Similar to the differential evolution algorithm in \cite{Double_protograph}, we use the differential evolution algorithm to find the protograph with the optimized BEC threshold. 
The parameters for the differential evolution are given as follows.
The number of generations of the algorithm $g$ is set to $6000$. Each entry of the base matrix can have the integer value varying from $0$ to a positive integer $t$.
The number of base matrices examined for each generation instance is defined as $N_p$. 
For a given base matrix size $n_c \times n_v$, we fix $N_p=10{\cdot}n_{c}n_{v}$. 
The mutation parameter $F$ is fixed to $0.5$ and $\alpha$ is a uniform random variable with the domain $[0,1]$. 
Lastly, the crossover probability $p_c$ is fixed to $0.88$ in this paper. \\
\indent We define the optimized PD-LDPC code ensemble as $C_1$ and the optimization algorithm is given in Alg.~3. 
It is clear that while the optimization process is the same as that of the protograph LDPC codes, the indices of the partial doping represented by $\mathcal{X}$ are included, which show the protograph VNs that are doped by GC nodes. 
{Although the indices of $\mathcal{X}$ can be arbitrarily selected for code constructions using the differential evolution algorithms, we fix the number of indices as small as possible. For applications on partially doping over a given protograph, algorithms selecting the indices of $\mathcal{X}$ can be made to optimize the performance of the code ensemble.}
Also, the criterion for the existence of the typical minimum distance derived in Theorem 1 is used during the construction of new trial matrices for the proposed PD-GLDPC codes. The component code used in the following optimization is a $(15,11)$ Hamming code.\\
\begin{figure*}[t!]
\begin{equation}
\small
\bold{B}_{8\times16}^{C_1}=
\left[
\label{diff_mat_0.5_rand}
\begin{array}{cccccccccccccccc}
\bold{5} & \bold{2} & 0 & 0 & 0 & 0 & 0 & 0 & 1 & 1 & 1 & 0 & 0 & 0 & 0 & 0\\
\bold{4} & \bold{0} & 0 & 0 & 0 & 0 & 0 & 0 & 0 & 1 & 0 & 1 & 0 & 1 & 0 & 0\\
\bold{5} & \bold{5} & 0 & 1 & 0 & 0 & 1 & 0 & 0 & 0 & 0 & 1 & 1 & 1 & 0 & 0\\
\bold{5} & \bold{0} & 1 & 0 & 2 & 1 & 5 & 5 & 0 & 0 & 5 & 5 & 0 & 5 & 0 & 3\\
\bold{5} & \bold{3} & 0 & 1 & 1 & 1 & 5 & 1 & 0 & 0 & 0 & 0 & 1 & 1 & 2 & 0\\
\bold{0} & \bold{0} & 1 & 0 & 0 & 0 & 0 & 1 & 4 & 1 & 3 & 0 & 0 & 2 & 1 & 0\\
\bold{4} & \bold{0} & 0 & 0 & 0 & 0 & 0 & 1 & 1 & 0 & 1 & 0 & 0 & 0 & 0 & 0\\
\bold{0} & \bold{5} & 0 & 0 & 0 & 0 & 1 & 0 & 3 & 0 & 0 & 1 & 1 & 1 & 1 & 0
\end{array}
\right]
\end{equation}
\end{figure*}
\begin{figure*}[t!]
\begin{equation}
\small
\bold{B}_{4\times12}^{C_1}=
\left[
\label{diff_mat_0.33_rand}
\begin{array}{cccccccccccc}
\bold{3} & 0 & 0 & 3 & 1 & 3 & 0 & 1 & 2 & 2 & 3 & 0\\
\bold{3} & 0 & 1 & 3 & 1 & 0 & 0 & 0 & 1 & 2 & 0 & 1\\
\bold{3} & 1 & 0 & 3 & 0 & 0 & 0 & 0 & 0 & 1 & 0 & 2\\
\bold{3} & 3 & 3 & 3 & 0 & 0 & 3 & 1 & 0 & 1 & 0 & 0
\end{array}
\right].
\end{equation}
\end{figure*}
\indent We optimize the protographs for the PD-GLDPC codes for base matrices with size $8\times16$ and $4 \times 12$.
We set $\mathcal{X}=\{1,2 \}$ and $t=5$ for $\bold{B}_{8\times16}$, $\mathcal{X}=\{1 \}$ and $t=3$ for $\bold{B}_{4\times12}$.
Let $\bold{B}_{n_c\times n_v}^{C_1}$ be the resulting base matrix of the optimization for both cases. The optimized base matrix result of $\bold{B}_{8\times16}^{C_1}$ is given in (\ref{diff_mat_0.5_rand}), where the BEC threshold is $0.5227$ with the code rate $0.4667$.
Likewise, the result of $\bold{B}_{4\times12}^{C_1}$ is given in (\ref{diff_mat_0.33_rand}), where the BEC threshold is $0.3397$ with the code rate $0.6444$.
The bold parts in the matrix represent VNs that are partially doped.
The results show that the VN with the highest degree is partially doped. 
From these optimization results, we can expect that partially doping VNs with high degree and puncturing some portion of them for rate matching can improve the performance of the proposed PD-GLDPC codes.\\
\indent {The approach of partially doping and puncturing is a similar techique to the precoding and puncturing. }
Precoding and puncturing high degree VNs in a protograph is a well known technique in order to enhance the threshold of protograph LDPC codes \cite{precoding_divsalar}.
Precoding takes place by placing a CN between a degree-1 VN and a high degree VN. 
In order to compensate for the rate loss, the high degree VN is punctured. 
From some intuition of the proposed optimization results and well known concepts of precoding, we apply a similar approach of the precoding technique to the proposed PD-GLDPC codes. \\
\indent We first define $\rho_d$ as the portion of random puncturing for VNs that are doped. 
For the BEC, we use the concept in \cite{randpunc} to derive $\rho_d$. For a target code rate $R^*$, the random puncturing ratio $\rho$ is $1-\frac{R}{R^*}$. 
Thus, $\rho_d$ is derived as $\rho_d=\rho \cdot \frac{n_v}{|\mathcal{X}|}$ and we use it for the computation of the  EXIT during the optimization algorithm.
The channel values for the partially doped VNs become $I_{ch}(j)=1-\{\rho_d + (1-\rho_d)\epsilon\}, \, j\in \mathcal{X}$. 
Thus, it is possible to construct the PD-GLDPC codes for the target code rate by using the random puncturing method. \\
\indent For the construction of PD-GLDPC codes with the target code rate $R^*=1/2$, the base matrix $\bold{B}_{8\times16}$ is optimized using Alg.~3 for $\mathcal{X}=\{ 1,2 \}$, $\rho_d=0.5333$, and $t=5$. 
Likewise, for the target code rate $R^*=2/3$, the base matrix $\bold{B}_{4\times12}$ is optimized for $\mathcal{X}=\{ 1 \}$, $\rho_d=0.4058$, and $t=3$.
Let $\bold{B}_{n_c\times n_v}^{C_2}$ be the resulting base matrix for the optimized results of the protographs constructed by puncturing partially doped VNs.
The optimized base matrices for both code rates are given as
\begin{figure*}[t]
\begin{equation}
\footnotesize
\bold{B}_{8\times16}^{C_2}=
\left[
\label{diff_mat_0.5_dopepunc}
\begin{array}{cccccccccccccccc}
\bold{2} & \bold{0} & 5 & 2 & 1 & 3 & 0 & 0 & 1 & 0 & 0 & 1 & 0 & 0 & 0 & 1\\
\bold{2} & \bold{0} & 0 & 0 & 1 & 0 & 0 & 0 & 0 & 0 & 0 & 1 & 1 & 0 & 0 & 0\\
\bold{1} & \bold{2} & 0 & 0 & 0 & 0 & 1 & 0 & 0 & 0 & 0 & 0 & 0 & 0 & 1 & 0\\
\bold{1} & \bold{0} & 0 & 0 & 0 & 0 & 0 & 1 & 0 & 2 & 1 & 0 & 2 & 1 & 2 & 0\\
\bold{2} & \bold{0} & 0 & 0 & 0 & 0 & 0 & 1 & 0 & 1 & 0 & 0 & 0 & 0 & 0 & 0\\
\bold{1} & \bold{1} & 1 & 0 & 1 & 0 & 0 & 0 & 1 & 0 & 0 & 0 & 2 & 1 & 1 & 0\\
\bold{0} & \bold{0} & 0 & 1 & 2 & 0 & 4 & 0 & 1 & 0 & 1 & 0 & 0 & 0 & 3 & 1\\
\bold{3} & \bold{1} & 1 & 0 & 1 & 0 & 0 & 0 & 0 & 0 & 0 & 1 & 0 & 0 & 0 & 0
\end{array}
\right],
\end{equation}
\end{figure*}
\begin{figure*}[t]
\begin{equation}
\footnotesize
\bold{B}_{4\times12}^{C_2}=
\left[
\label{diff_mat_0.33_dopepunc}
\begin{array}{cccccccccccc}
\bold{2} & 0 & 0 & 1 & 0 & 0 & 3 & 2 & 1 & 0 & 2 & 2\\
\bold{2} & 0 & 1 & 1 & 0 & 0 & 1 & 1 & 0 & 0 & 2 & 2\\
\bold{3} & 1 & 0 & 0 & 0 & 1 & 1 & 0 & 0 & 0 & 0 & 1\\
\bold{2} & 1 & 2 & 0 & 3 & 2 & 0 & 0 & 1 & 3 & 0 & 3
\end{array}
\right],
\end{equation}
\end{figure*}
where resulting base matrix for $R^*=1/2$ is given in (\ref{diff_mat_0.5_dopepunc}) and the resulting base matrix for $R^*=2/3$ is given in (\ref{diff_mat_0.33_dopepunc}). 
The resulting thresholds of the optimized base matrices are $0.4857$ and $0.319$ for target code rates $R^*=1/2$ and $R^*=2/3$, respectively. 
The optimization results show that the constructed PD-GLDPC codes have capacity approaching performances and the average VN density is reduced by huge amount compared to $\bold{B}_{n_c\times n_v}^{C_1}$. 
Since the base matrix $\bold{B}_{n_c\times n_v}^{C_2}$ is driven from the random puncturing of partially doped VNs, we define the constructed PD-GLDPC code ensemble with parameters $(\bold{B}_{n_c\times n_v}^{C_2},\mu,\kappa,\mathcal{X},\rho_d)$.

\vspace{10pt}
\section{Numerical Results and Analysis}
In this section, we propose the optimized protograph design and show the BLER of the proposed protograph-based PD-GLDPC codes.
The performance of the conventional protograph LDPC code is compared with that of the proposed protograph-based PD-GLDPC code. In order to make this comparison, we construct a protograph ensemble that has the same variable node degree distribution as that of the proposed protograph-based PD-GLDPC code.

In this section, we propose the optimized protograph design and show the FER of the proposed PD-GLDPC codes.
The performance of the conventional protograph LDPC code is compared with that of the proposed PD-GLDPC code.
Two methods of comparison are conducted. The first subsection compares them under the same degree distribution using Alg.~2. 
The second subsection compares the performance of the PD-GLDPC codes constructed without the degree distribution constraints using Alg.~3 to the state-of-the-art protograph LDPC codes.

\subsection{Simulation Result for Optimized PD-GLDPC Code from Irregular Random LDPC Code Ensembles}
\begin{figure}
    \centering
    \includegraphics[draft=false,width=8cm]{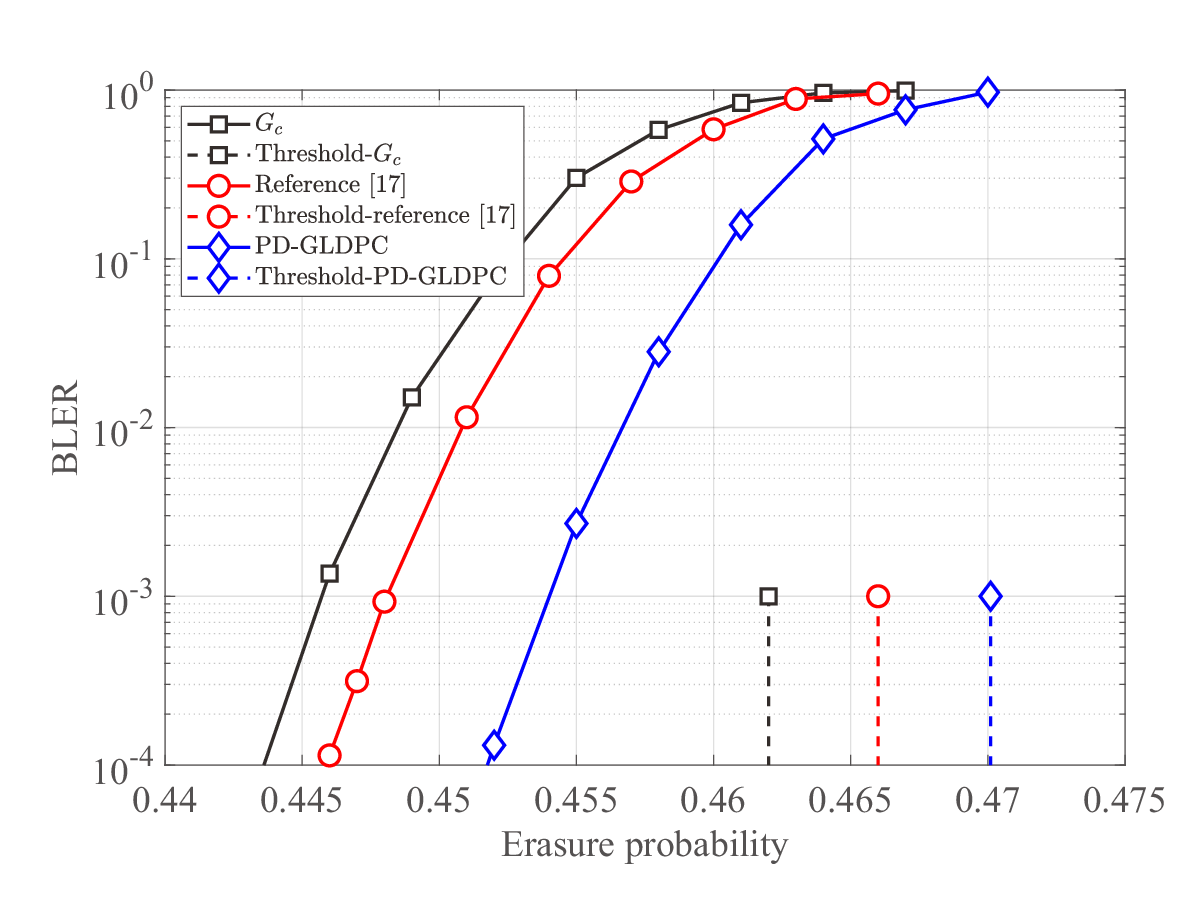}
    \caption{
   { Comparison of the BEC threshold and FER for the LDPC codes constructed from AR4JA and $G_c$, the conventional random GLDPC code from the ensemble in} \cite{random_GLDPC}, {and the PD-GLDPC code from $G_p$ for the code rate $1/2$.}}
    \label{fig:IRR_half}
\end{figure}

As the performance comparison with the existing GLDPC codes, we use the random GLDPC code ensemble with the threshold $0.466$ in \cite{random_GLDPC} that is represented as  $\lambda(x)=0.8x^2+0.01x^3+0.01x^5+0.18x^7$ and {{a doping ratio $\nu=0.4$}} by the Hamming code. 
Fig.~\ref{fig:IRR_half} shows performance comparison of four half-rate codes which are AR4JA code \cite{div_proto}, the irregular protograph LDPC code constructed from $G_c$, the random ensemble-based GLDPC code in \cite{random_GLDPC}, and the proposed PD-GLDPC code constructed from $G_p$ in Table~2, where $y_{max}=5$. 
All four codes in Fig.~\ref{fig:IRR_half} are $(n,k)=(30000,15000)$ codes of the  half-rate, where $G_c$ is defined as $\bold{D}^{(2,3,4,5,6,20)}=(165,134,47,23,5,26)$ and has the same VN degree distribution as the PD-GLDPC code after lifting by $N=75$. $x=\mu y=75$ protograph VNs are partially doped in the PD-GLDPC code. {For the constructed PD-GLDPC code, we have $\nu=\frac{x\beta }{x\beta+n_c N}=\frac{375}{375+15000}=0.02439$}.
The constructed PD-GLDPC code for $y_{max}=5$ has a coding gain of 0.0079 and 0.0039 compared to the GLDPC code in \cite{random_GLDPC} and the irregular protograph LDPC code from $G_c$, respectively. 
Fig.~\ref{fig:IRR_half} shows that the proposed PD-GLDPC code has a good performance both in the waterfall and the low error floor region due to the fact that the code is optimized by increasing the doping as much as possible, and at the same time, the typical minimum distance constraint is satisfied. 
In terms of the asymptotic analysis, increasing the portion of degree-2 VNs increases the possibility of the code to approach the channel capacity \cite{deg2_thres_IT}.
However, the existence of a typical minimum distance of the protograph is also important, which upper bounds the portion of degree-2 VNs in the LDPC code. 
Thus, balancing the portion of degree-2 VNs is needed in order to satisfy both a typical minimum distance condition and a good threshold. 
The proposed PD-GLDPC code guarantees the balance of the degree-2 VNs by carefully choosing the rate of the protograph code and the number of doping on degree-2 VNs. 

\subsection{Simulation Results for PD-GLDPC Code from Optimized Protograph}

\begin{figure}
    \centering
    \subfloat[$R^*=\frac{1}{2}$]{\includegraphics[draft=false,width=8cm]{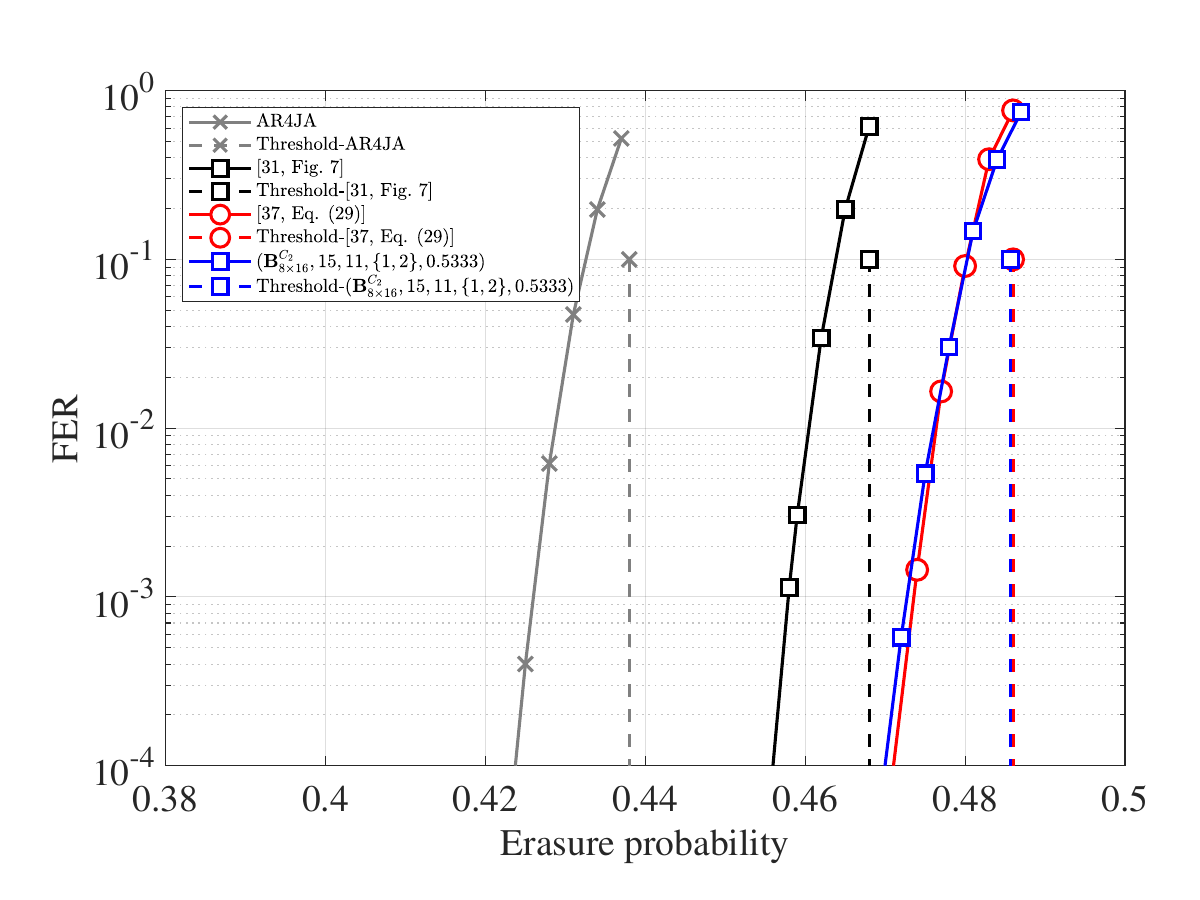}}
    \subfloat[$R^*=\frac{2}{3}$]{\includegraphics[draft=false,width=8cm]{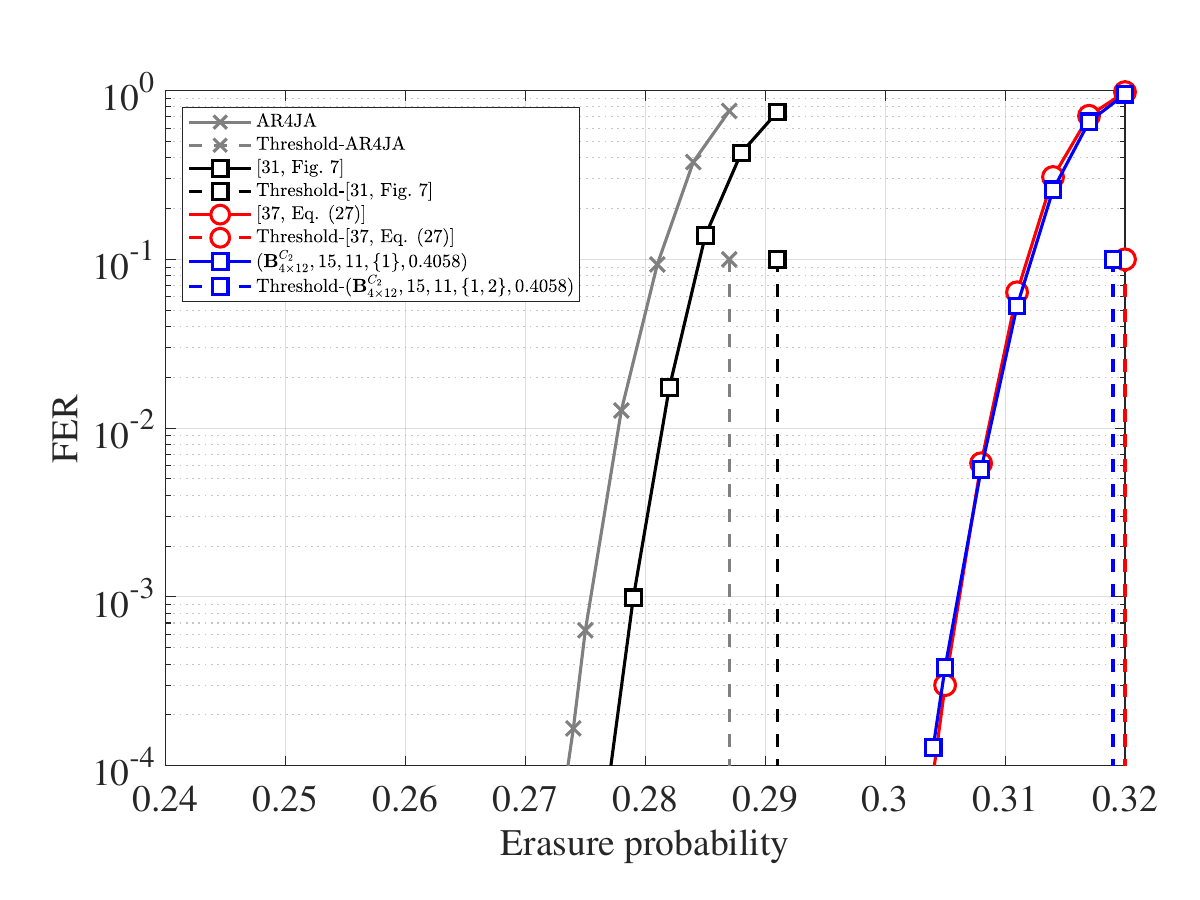}}
    
    \caption{{FER comparison for the constructed codes from AR4JA}~\cite{div_proto}, {protograph}~[31, Fig.~7], {protograph}~\cite{Double_protograph}, {and PD-GLDPC code ensemble $(\bold{B}_{n_c\times n_v}^{C_2},15,11,\mathcal{X},\rho_d)$}.
    }
\label{fig:diff_result}
\end{figure}

\begin{table*}[t]
\caption{{Comparison for thresholds and average VN degrees of protographs for the BEC.}}
\tiny
\resizebox{\textwidth}{!}{
\begin{tabular}{|c|c|c|c|c|c|}
\hline
\textbf{Code type}                                         & \textbf{Code rate} & \textbf{Protograph size} & \textbf{Threshold} & \textbf{Average VN degree} & \textbf{Gap to capacity} \\ \hline
AR4JA~\cite{div_proto}              & 0.5             & 3$\times$5               & 0.438            & 3                        & 0.062                  \\ \hline
Protograph [31, Fig.~7]                      & 0.5              & 4$\times$8               & 0.468            & 4.25                     & 0.032                 \\ \hline
Protograph~\cite{Double_protograph} & 0.5              & 8$\times$16              & 0.486            & 5.25                     & 0.014                  \\ \hline
PD-GLDPC                                                   & 0.5              & 8$\times$16              & 0.4857           & 4.58                     & 0.0143                \\ \hline\hline
AR4JA \cite{div_proto}              & 0.67             & 3$\times$7               & 0.287            & 3.29                     & 0.046                  \\ \hline
Protograph~{[}31, Fig.~7{]}                      & 0.67             & 2$\times$6               & 0.292            & 5                        & 0.041                  \\ \hline
Protograph \cite{Double_protograph} & 0.67             & 4$\times$12              & 0.320            & 5.08                     & 0.013                  \\ \hline
PD-GLDPC                                                   & 0.67             & 4$\times$12              & 0.319            & 4.09                     & 0.014                  \\ \hline
\end{tabular}
}
\end{table*}

\indent The proposed PD-GLDPC codes for $R^*=1/2$ and $R^*=2/3$ are constructed from the ensembles $(\bold{B}_{8\times 16}^{C_2},15,11,\{1,2 \},0.5333)$ and $(\bold{B}_{4\times 12}^{C_2},15,11,\{1 \},0.4058)$, respectively. 
The protographs are shown in  (\ref{diff_mat_0.5_dopepunc}) and (\ref{diff_mat_0.33_dopepunc}). 
The AR4JA \cite{div_proto} and block protograph codes in \cite{div_proto,Double_protograph} of the same code rate are used for performance comparison. 
We first compare the threshold and average VN degree between the proposed PD-GLDPC code ensembles and the aforementioned protograph LDPC code ensembles in Table~3. 
The average VN degree of the PD-GLDPC codes considers both the base matrix and the edges added from the partial doping.
The results show that the asymptotic performance of the proposed PD-GLDPC code ensemble outperforms the AR4JA and block protograph introduced in~\cite{div_proto}. 
The average VN degree of the PD-GLDPC codes is low while having the asymptotic performance comparable to the capacity approaching protographs introduced in \cite{Double_protograph}.\\
\indent By using the PEG algorithm, the protographs are lifted to construct $(48000,24000)$ PD-GLDPC code for $R^*=1/2$. The protograph AR4JA and protographs in \cite{div_proto,Double_protograph} are lifted to the same code length. The FER results are shown in Fig.~\ref{fig:diff_result}(a). Likewise, the protograph of the PD-GLDPC, AR4JA, and \cite{Double_protograph} are lifted to construct $(45000,30000)$ codes for $R^*=2/3$. The protograph in [31, Fig.~7] is lifted to blocklength near $n=45000$.
The FER results are shown in Fig.~\ref{fig:diff_result}(b). 
{{The doping ratio $\nu$ for the PD-GLDPC codes is $0.016393$ for both code rates $R^*=1/2$ and $R^*=2/3$.
Also, the FER results of the proposed PD-GLDPC codes for both code rates $R^*=1/2$ and $R^*=2/3$ show tangible gain compared to the AR4JA code and protograph code in}}~\cite{div_proto}. 
Also, the performance is comparable to the capacity approaching block LDPC code in~\cite{Double_protograph}. 
The partial doping and puncturing technique, which is similar to the precoding technique, shows that the capacity approaching PD-GLDPC codes can be constructed with the relatively low average VN degree. 

\vspace{10pt}
\section{Conclusion}
We proposed a new class of GLDPC codes called PD-GLDPC codes that has advantages of a finer doping granularity compared to the conventional protograph doped GLDPC codes.
Also, we proposed two optimization algorithms for the PD-GLDPC codes: protographs constructed from random LDPC code ensembles and protographs for PD-GLDPC code ensembles constructed from genetic algorithms. 
Furthermore, we proposed the partially doping and puncturing technique. 
Using the proposed technique, the constructed PD-GLDPC codes have good FER performances compared to the popular protograph LDPC codes. 
Since it is possible to partially dope the protograph VNs with a granularity one, the rate loss is reduced from partial doping, and thus, GLDPC codes can have capacity approaching performance in the medium to high code rate regime.
{For future work, use of other component codes and protographs with degree-1 VNs can be studied. Also, constructions of PD-GLDPC codes by generalizing the partial doping process such as doping over multiple protograph VNs or doping only a portion of a protograph VN can be considered. Furthermore, new constructions of PD-GLDPC codes over additive white Gaussian noise channels can be made.}

\section*{{Abbreviations}}
\vspace{10pt}
\begin{tabular}{l@{$\hspace{25pt}$}p{12cm}}
\textbf{AR4JA} & Accumulate-Repeat-4-Jagged-Accumulate \\
\textbf{BEC}  & Binary erasure channel \\
\textbf{CN} & Check node \\
\textbf{EXIT} & Extrinsic information transfer \\
\textbf{FER} & Frame error rate \\
\textbf{GC} & Generalized constraint\\
\textbf{GLDPC} & Generalized low-density parity-check\\
\textbf{LDPC} & Low-density parity-check\\
\textbf{ML} & Maximum likelihood\\
\textbf{PCM} & Parity-check matrix\\
\textbf{PD-GLDPC} & Partially doped GLDPC\\
\textbf{PEG} & Progressive edge growth\\
\textbf{PEXIT} & Protograph EXIT\\
\textbf{SPC} & Single parity-check\\
\textbf{VN} & Variable node\\
\end{tabular}

\vspace{10pt}
\appendices
\section{Proof of Theorem 1: The Constraint for the Existence of the Typical Minimum Distance of the Proposed Protograph-based PD-GLDPC Codes}
A proof for the constraint of the existence of a typical minimum distance for the proposed protograph-based PD-GLDPC codes is given in this appendix.
Similar to that in~\cite{Tdmin_GLDPC}, a typical minimum distance is driven by the weight enumerator analysis over the lifted protograph. 
In order to use the notations in \cite{Tdmin_GLDPC}, we've distinguished the indexing notations during the enumeration for the partially doped variable nodes using $'$. 
Also, the $c_j$ and $v_i$ notations are used for the check nodes and the variable nodes, respectively. 
Suppose that the proposed protograph-based PD-GLDPC code is constructed from the protograph defined by $G=(V,C,E)$ and the $x$ variable nodes are partially doped, where component codes are identical with the parameters $(\mu,\kappa)$.
We are given a variable node set $V=\{ v_1,{\cdots},v_{n_v} \}$ and a check node set $C_{PD{\text{-}}GLDPC}= B_{Hamm} \cup C =  \{b_1,{\cdots},b_{x} \} \cup \{c_1,{\cdots},c_{n_c} \}$ for the protograph.
It is important to note that the GC node set $B_{Hamm}$ is not defined over a protograph.
However, the codeword enumeration can be made when the protograph is lifted, where $b_{i'},i'\in [x]$ is a virtual check node that represents the Hamming check nodes used for partial doping for $v_{i'}$ in the original protograph. 
Although $b_{i'}$ is not a protograph check node, we define it for the enumeration of the partially doped protograph variable nodes. 
The protograph-based PD-GLDPC code is constructed by lifting the graph $G$ by $N$ times and permuting the replicated edges. 
Each $v_i$ ($c_j$) has degree $q_{v_i}$ ($q_{c_j}$) and each $b_{i'}$ has degree $\mu$. 
For the enumeration of the GC node $b_{i'}$, we can think of it as a protograph node of degree $\mu$ that is lifted by a factor of $\frac{N}{\mu}$. 
The upper bound of the weight enumerator of the proposed protograph-based PD-GLDPC code with weight $d$, denoted as $A_{d}^{PD{\text{-}}GLDPC}$ is derived as follows.\\
\indent Let $w_{m,u},u \in [q_{v_m}]$ be the $u$th edge weight from a variable node $v_m$. 
For a partially doped variable node $v_m, m\in [ x]$, there are $\mu$ weights sent towards the incident GC node, where the $u$th weight is defined as $w'_{m,u}, u\in [\mu]$. 
For a given input weight vector $\bold{d}=(d_1,{\cdots},d_{n_v} )$, we need to calculate $A^{PD{\text{-}}GLDPC}(\bold{d})$ and sum it over every instance of $\bold{d}$ that satisfies $d=d_1+{\cdots}+d_{n_v}$. For input $d_{i'},i'\in [x]$, it is clear that $\sum_{i=1}^{\mu}w'_{m,i}=d_{i'}$ because the extrinsic weight $w'_{m,i}$ consists of weights solely from $v_m$. 
We introduce the following notations: 
\begin{itemize}
    \item $A^{v_i}_{d_i}(\bold{w}_i)=\binom{N}{d_i}\delta_{d_i,w_{i,1}},{\cdots},\delta_{d_i,w_{i,q_{v_i}}}= \begin{cases} \binom{N}{d_i},  \quad \mbox{if} \, w_{i,j}=d_i,\forall j\in [q_{v_i}] \\ 0, \quad \mbox{otherwise} \end{cases}$ is the vector weight enumerator for a variable node $v_i$ of the protograph \cite{Tdmin_GLDPC}. 
    \item $A^{c_j}(\bold{z}_j)$ is the vector weight enumerator for a check node $c_j$ of the original protograph, for the incoming weight vector $\bold{z}_j = [z_{j,1},{\cdots},z_{j,q_{c_j}}]$ \cite{Tdmin_GLDPC}.
    \item $B_{d_{i'}}^{v_{i'}}(\bold{w}'_{i'})=\begin{cases} 1, \quad \mbox{for}\, w'_{i',1}+{\cdots}+w'_{i',\mu}=d_{i'} \\ 0, \quad \mbox{otherwise} \end{cases}$ is the vector weight enumerator for partially doped variable nodes $v_{i'},\,i'\in [x]$.
    \item $B^{b_{i'}}(\bold{w}'_{i'})$ is the vector weight enumerator for check nodes that are created during the lifting process given the weight vector $\bold{w}'_{i'}$. $A^{b_{i'}}(d_{i'})$ is the summation of enumerators over all possible $\bold{w}'_{i'}$ values given that $w'_{i',1}+{\cdots}+w'_{i',\mu}=d_{i'}$ satisfying
\begin{align}
    A^{b_{i'}}(d_{i'})=\sum_{\bold{w}'}B^{b_{i'}}(\bold{w}'_{i'})=\sum_{\bold{w}'}\sum_{\{ \bold{m} \}}C(\frac{N}{\mu};m_1,{\cdots},m_K), \nonumber 
     \end{align}
where $\bold{w}'=( w_{1}',{\cdots},w_{\mu}' ) $ such that $\sum_{i=1}^{\mu}w_{1}'=d_{k}', \, w_{i}' \leq \frac{N}{\mu}.$
   
\end{itemize}
Then, the weight enumerator is given as 
\begin{align}
    A_{d}^{PD-GLDPC}=\sum_{ \{ \bold{d} \}}A^{PD-GLDPC}(\bold{d}),\nonumber
\end{align}
where \begin{align}
    A^{PD{\text{-}}GLDPC}(\bold{d})&=\frac{\prod_{i=1}^{n_v}A^{v_i}_{d_i}(\bold{w}_i)\prod_{j=1}^{n_c}A^{c_j}(\bold{z}_j)\times\prod_{i'=1}^{x}B_{d_{i'}}^{v_{i'}}(\bold{w}'_{i'})B^{b_{i'}}(\bold{w}'_{i'})}{\prod_{s=1}^{n_v}\prod_{r=1}^{q_{v_s}}\binom{N}{w_{s,r}}\times \prod_{s'=1}^{x}\prod_{r'=1}^{\mu}\binom{\frac{N}{\mu}}{w'_{s',r'}}}  \nonumber\\
    &=\sum_{\{\bold{w}'_{i'}: \, w'_{i',1}+{\cdots}+w'_{i',\mu}=d_{i'}\}} \frac{\prod_{j=1}^{n_c}A^{c_j}(\bold{d}_j)\times\prod_{i'=1}^{x}B^{b_{i'}}(\bold{w}'_{i'})}{\prod_{i=1}^{n_v}\binom{N}{d_i}^{q_{v_i}-1}\times \prod_{s'=1}^{x}\prod_{r'=1}^{\mu}\binom{\frac{N}{\mu}}{w'_{s',r'}}}.\nonumber
\end{align}

The solution to the equation $\bold{w}'=\bold{m}\bold{M^C}$ is given as $\bold{m}=\{m_1,{\cdots},m_K \}$. The term $\binom{\frac{N}{\mu}}{w'_{s',r'}}$ is lower bounded by $(\frac{N}{\mu})^{w'_{s',r'}}e^{-w'_{s',r'} \cdot {\rm{ln}} \, w'_{s',r'}}$. Then, $A^{PD-GLDPC}(\bold{d})$ can be upper bounded as

\begin{align}
     A^{PD{\text{-}}GLDPC}(\bold{d}) &\leq  \sum_{\{\bold{w}'_{i'}: \, w'_{i',1}+{\cdots}+w'_{i',\mu}=d_{i'}\}} \frac{\prod_{j=1}^{n_c}A^{c_j}(\bold{d}_j)\times\prod_{i'=1}^{x}B^{b_{i'}}(\bold{w}'_{i'})}{\prod_{i=1}^{n_v}\binom{N}{d_i}^{q_{v_i}-1}\times \prod_{s'=1}^{x}\prod_{r'=1}^{\mu}(\frac{N}{\mu})^{w'_{s',r'}}e^{-w'_{s',r'} \cdot {\rm{ln}} \, w'_{s',r'}}} \nonumber \\
     &\leq \sum_{\{\bold{w}'_{i'}: \, w'_{i',1}+{\cdots}+w'_{i',\mu}=d_{i'}\}} \frac{\prod_{j=1}^{n_c}A^{c_j}(\bold{d}_j)\times\prod_{i'=1}^{x}B^{b_{i'}}(\bold{w}'_{i'})}{\prod_{i=1}^{n_v}\binom{N}{d_i}^{q_{v_i}-1}\times \prod_{s'=1}^{x}(\frac{N}{\mu})^{d_{s'}}e^{-d_{s'} \cdot {\rm{ln}} \, d_{s'}}} \nonumber \\
     &\leq \sum_{\{\bold{w}'_{i'}: \, w'_{i',1}+{\cdots}+w'_{i',\mu}=d_{i'}\}} \frac{\prod_{j=1}^{n_c}A^{c_j}(\bold{d}_j)\times\prod_{i'=1}^{x}B^{b_{i'}}(\bold{w}'_{i'})}{\prod_{i=1}^{n_v}\binom{N}{d_i}^{q_{v_i}-1}\times (\frac{N}{\mu})^{P}e^{-P \cdot {\rm{ln}} \, P}} \nonumber \\                           
     &=\frac{\prod_{j=1}^{n_c}A^{c_j}(\bold{d}_j)\times \prod_{i'=1}^{x} \sum_{\{\bold{w}'_{i'}: \, w'_{i',1}+{\cdots}+w'_{i',\mu}=d_{i'} \}}B^{b_{i'}}(\bold{w}'_{i'})}{\prod_{i=1}^{n_v}\binom{N}{d_i}^{q_{v_i}-1} \times (\frac{N}{\mu})^{P}e^{-P \cdot {\rm{ln}} \, P}} \nonumber \\
     &=\frac{\prod_{j=1}^{n_c}A^{c_j}(\bold{d}_j)\times \prod_{i'=1}^{x}A^{b_{i'}}(d_{i'})}{\prod_{i=1}^{n_v}\binom{N}{d_i}^{q_{v_i}-1} \times (\frac{N}{\mu})^{P}e^{-P \cdot {\rm{ln}} \, P}},
     \label{eq:GLD_P}
\end{align}
where $P=\sum_{s'=1}^{x}d_{s'}$ is the total weight of the $x$ partially doped variable nodes.  Then
$\sum_{t}(t \cdot {\rm{ln}}\,t) \leq (\sum_{t}t)\cdot {\rm{ln}}\, (\sum_{t}t)$ is used for the second and the third inequalities in~(\ref{eq:GLD_P}). 
It was shown in~(18) of \cite{Tdmin_GLDPC} that the inequality $$\frac{\prod_{j=1}^{n_c}A^{c_j}(\bold{d}_j)}{\prod_{i=1}^{n_v}\binom{N}{d_i}^{q_{v_i}-1}}\leq \prod_{i=1}^{n_v}e^{(q_{v_{i}}-1-\frac{q_{v_{i}}}{d_{min}^{(c)}})d_i {\rm{ln}}\,\frac{d_i}{N}+\frac{q_{v_{i}}(2+k_{max}^{(c)}{\rm{ln}\,2)}}{d_{min}^{(c)}}d_i}$$ holds, where $d_{min}^{(c)}$ is the minimum distance of an SPC component code for the original protograph and $k_{max}^{(c)}$ is the maximum number of codewords of an SPC component code. 
Using the similar notations in~\cite{Tdmin_GLDPC}, let $d_{min}^{(b)}$ and $k^{(b)}$ be the minimum distance and the number of codewords of the $(\mu,\kappa)$ component code for the GC nodes.
Then, $\prod_{i'=1}^{x}A^{b_{i'}}(d_{i'})$ is upper bounded as
\begin{align}
    \prod_{i'=1}^{x}A^{b_{i'}}(d_{i'}) &\leq \prod_{i'=1}^{x}\sum_{\{\bold{w}'_{i'}: \, w'_{i',1}+{\cdots}+w'_{i',\mu}=d_{i'} \}}\prod_{i=1}^{\mu}{(\frac{N}{\mu})}^{\frac{1}{d_{min}^{(b)}}  w'_{i',i}}e^{\frac{(2+k'_{i'}ln \, 2)}{d_{min}^{(b)}}w'_{i',i} - \frac{1}{d_{min}^{(b)}}w'_{i',i} {\rm{ln}} \, w'_{i',i} } \nonumber \\
    &=\prod_{i'=1}^{x}\sum_{\{\bold{w}'_{i'}: \, w'_{i',1}+{\cdots}+w'_{i',\mu}=d_{i'} \}}(\frac{N}{\mu})^{\frac{1}{d_{min}^{(b)}}d_{i'}}e^{\frac{(2+k'_{i'}{\rm{ln}} \, 2)}{d_{min}^{(b)}}d_{i'}-\sum_{i=1}^{\mu}\frac{1}{d_{min}^{(b)}}w'_{i',i} {\rm{ln}} \, w'_{i',i}} \nonumber \\
    &\leq \prod_{i'=1}^{x}\sum_{\{\bold{w}'_{i'}: \, w'_{i',1}+{\cdots}+w'_{i',\mu}=d_{i'} \}}(\frac{N}{\mu})^{\frac{1}{d_{min}^{(b)}}d_{i'}}e^{\frac{(2+k'_{i'}{\rm{ln}} \, 2)}{d_{min}^{(b)}}d_{i'}-\frac{1}{d_{min}^{(b)}}d_{i'} {\rm{ln}} \, \frac{d_{i'}}{\mu}} \nonumber \\
     &\leq \prod_{i'=1}^{x}\binom{d_{i'}+\mu-1}{d_{i'}}(\frac{N}{\mu})^{\frac{1}{d_{min}^{(b)}}d_{i'}}e^{\frac{(2+k'_{i'}{\rm{ln}} \, 2)}{d_{min}^{(b)}}d_{i'}-\frac{1}{d_{min}^{(b)}}d_{i'} {\rm{ln}} \, \frac{d_{i'}}{\mu}}.
    \label{eq:di+upper}
\end{align}
For the inequality in the third line of~(\ref{eq:di+upper}), we use the fact that $\sum_{i=1}^{p}t_i {\rm{ln}}\, t_i \leq s \cdot {\rm{ln}}\, \frac{s}{p}$ with $s=t_1+{\cdots}+t_p$, which is clear by using the derivative on the multivariable function that consists of independent $t_i$'s. 
The equality is satisfied when all $t_i$ values are the same. 
Going back to~(\ref{eq:GLD_P}), let $f(P)=\frac{1}{\binom{N}{\mu}^P e^{-P {\rm{ln}}\, P}}$ for convenience. Then we have
\begin{align}
    A^{PD{\text{-}}GLDPC}(\bold{d}) \leq& \prod_{i=1}^{n_v}e^{(q_{v_{i}}-1-\frac{q_{v_{i}}}{d_{min}^{(c)}})d_i {\rm{ln}}\,\frac{d_i}{N}+\frac{q_{v_{i}}(2+k_{max}^{(c)}{\rm{ln}\,2)}}{d_{min}^{(c)}}d_i} \nonumber \\ &\times \prod_{i'=1}^{x}e^{d_{i'}+\mu-1}(\frac{N}{\mu})^{\frac{1}{d_{min}^{(b)}}d_{i'}}e^{\frac{(2+k'_{i'}{\rm{ln}} \, 2)}{d_{min}^{(b)}}d_{i'}-\frac{1}{d_{min}^{(b)}}d_{i'} {\rm{ln}} \, \frac{d_{i'}}{\mu}} \times f(P) \nonumber \\ 
    \leq& \prod_{i=1}^{n_v}e^{(q_{v_{i}}-1-\frac{q_{v_{i}}}{d_{min}^{(c)}})d_i {\rm{ln}}\,\frac{d_i}{N}+\frac{q_{v_{i}}(2+k_{max}^{(c)}{\rm{ln}\,2)}}{d_{min}^{(c)}}d_i} \nonumber \\ &\times e^{x(\mu-1)}e^P\prod_{i'=1}^{x}(\frac{N}{\mu})^{\frac{1}{d_{min}^{(b)}}d_{i'}}e^{\frac{(2+k^{(b)}{\rm{ln}} \, 2)}{d_{min}^{(b)}}d_{i'}-\frac{1}{d_{min}^{(b)}}d_{i'} {\rm{ln}} \, \frac{d_{i'}}{\mu}} \times f(P) .
    \label{eq:fP}
\end{align}
We classify the variable nodes in the protograph into three groups before doping:
\begin{itemize}
    \item Protograph variable nodes of degrees higher than 2
    \item Protograph variable nodes of degree 2 to be partially doped
    \item Protograph other variable nodes of degree 2.
\end{itemize}
We also separate the weights of codewords after lifting into three parts according to the three groups of variable nodes: $u_i$, $p_z$, and $l_j$, where $u_i$ is the weight of the sub-codeword corresponding to a protograph variable node $v_i$ of degree higher than 2 and $p_z$ and $l_j$ are the weights of codewords of each partially doped and undoped protograph variable node $v_z$ and $v_j$ of degree 2 from the protograph, respectively.
The sum of sub-codeword weights for each group of variable nodes is given as $U=\sum_{i}u_i$, $P=\sum_{z}p_z$, and $L=\sum_{j}l_j$. 
It is clear that for the total codeword weight $d$, $d=U+P+L$. 
Then, the upper bound of the first term in~(\ref{eq:fP}) is written as
\begin{align}
    \prod_{i=1}^{n_v}e^{(q_{v_{i}}-1-\frac{q_{v_{i}}}{d_{min}^{(c)}})d_i {\rm{ln}}\,\frac{d_i}{N}+\frac{q_{v_{i}}(2+k_{max}^{(c)}{\rm{ln}\,2)}}{d_{min}^{(c)}}d_i}\leq &e^{(2-\frac{3}{d_{min}^{(c)}})(d-P-L){\rm{ln}}\,\frac{d-P-L}{N}+\frac{3(2+k_{max}^{(c)})}{d_{min}^{(c)}}\cdot (d-P-L)} \nonumber \\
    &\times e^{\frac{2(2+k_{max}^{(c)})}{d_{min}^{(c)}}\cdot L}\cdot e^{\frac{2(2+k_{max}^{(c)})}{d_{min}^{(c)}}\cdot P},
    \label{eq:GLDPC_enum}
\end{align}
which is derived by using three weight groups of codewords similar to~(20) of \cite{Tdmin_GLDPC}. We share the same inequality $u_i < Ne^{-\frac{(2+k_{max}^{(c)}{\rm{ln}}\,2)}{d_{min}^{(c)}-1}}$ over the given codeword weight $d$ as in \cite{Tdmin_GLDPC}.
Using the derivation in~\cite{Tdmin_GLDPC}, the upper bound of the second term of~(\ref{eq:fP}) can be derived as
\begin{align}
    &\prod_{i'=1}^{x}(\frac{N}{\mu})^{\frac{1}{d_{min}^{(b)}}d_{i'}}e^{\frac{(2+k^{(b)}{\rm{ln}} \, 2)}{d_{min}^{(b)}}d_{i'}-\frac{1}{d_{min}^{(b)}}d_{i'} {\rm{ln}} \, \frac{d_{i'}}{\mu}}=\prod_{i'=1}^{x}e^{\frac{1}{d_{min}^{(b)}}d_{i'}{\rm{ln}} \, \frac{N}{\mu}}e^{\frac{(2+k^{(b)}{\rm{ln}} \, 2)}{d_{min}^{(b)}}d_{i'}-\frac{1}{d_{min}^{(b)}}d_{i'} {\rm{ln}} \, \frac{d_{i'}}{\mu}} \nonumber \\
    &=\prod_{i'=1}^{x}e^{\frac{1}{d_{min}^{(b)}}d_{i'}{\rm{ln}} \, \frac{N}{d_{i'}}}e^{\frac{(2+k^{(b)}{\rm{ln}} \, 2)}{d_{min}^{(b)}} d_{i'}}=\prod_{z=1}^{x}e^{\frac{1}{d_{min}^{(b)}}p_z{\rm{ln}} \, \frac{N}{p_z}}e^{\frac{(2+k^{(b)}{\rm{ln}} \, 2)}{d_{min}^{(b)}} p_z} \nonumber \\
    &\leq e^{\frac{1}{d_{min}^{(b)}}P\cdot {\rm{ln}}\,\frac{Nx}{P} }e^{\frac{(2+k^{(b)}{\rm{ln}} \, 2)}{d_{min}^{(b)}} P}.
    \label{eq:doped_upper}
\end{align}

Using~(\ref{eq:GLDPC_enum}) and~(\ref{eq:doped_upper}), the upper bound of $A^{PD{\text{-}}GLDPC}(\bold{d})$ is derived in terms of $E(d,P,L)$ as follows:
\begin{align}
    A^{PD{\text{-}}GLDPC}(\bold{d})\leq& e^{\frac{1}{d_{min}^{(b)}}P\cdot {\rm{ln}}\,\frac{Nx}{P} }e^{\frac{(2+k^{(b)}{\rm{ln}} \, 2)}{d_{min}^{(b)}} P} \cdot e^{(2-\frac{3}{d_{min}^{(c)}})(d-P-L){\rm{ln}}\,\frac{d-P-L}{N}+\frac{3(2+k_{max}^{(c)})}{d_{min}^{(c)}}\cdot (d-P-L)} \nonumber \\
    &\times e^{\frac{2(2+k_{max}^{(c)})}{d_{min}^{(c)}}\cdot (P+L)}\cdot e^{x(\mu-1)}e^P\cdot f(P).
    \label{eq:GLD_final_upper}
\end{align}
Let $E(d,P,L)$ be the parameter satisfying $A^{PD{\text{-}}GLDPC}(\bold{d})\leq e^{x(\mu-1)}\cdot e^{E(d,P,L)}$. Then, from the upper bound in (\ref{eq:GLD_final_upper}), $E(d,P,L)$ is given as 
\begin{align}
    E(d,P,L)=&\frac{1}{d_{min}^{(b)}}P\cdot {\rm{ln}}\,\frac{Nx}{P} +\frac{(2+k^{(b)}{\rm{ln}} \, 2)}{d_{min}^{(b)}} P + (2-\frac{3}{d_{min}^{(c)}})(d-P-L){\rm{ln}}\,\frac{d-P-L}{N} \nonumber \\ &+\frac{3(2+k_{max}^{(c)})}{d_{min}^{(c)}}\cdot (d-P-L)+\frac{2(2+k_{max}^{(c)})}{d_{min}^{(c)}}\cdot (P+L)+P+P{\rm{ln}}\,P-P{\rm{ln}}\, \frac{N}{\mu}. \nonumber
\end{align}

    Assuming that there are no type 1 degree-2 variable nodes defined in \cite{Tdmin_GLDPC} and there are no cycles consisting only of undoped variable nodes of degree-2, we use the result of~(22) in \cite{Tdmin_GLDPC} such that the inequality $l_{2,k}^{(c_j)}\leq \frac{1}{d_{min}^{(c_j)}}( L_2^{(c_j)} + \sum_{i}w_i^{(c_j)} )$ is satisfied for all $j\in[n_c]$, where $l_{2,k}^{(c_j)}$ is the weight of the degree-2 undoped variable node of $G_p$ and the total weight of them is denoted as $L_2^{(c_j)}$ for check node $c_j$. 
    Similar to the result in \cite{Tdmin_GLDPC}, we can derive the upper bound $L\leq \gamma(U+P)$, which is the same as $L \leq \frac{\gamma}{1+\gamma}d$. 
    Now, the upper bound of $E(d,P,L)$ needs to be derived for independent values $L$ and $P$. The first and second partial derivatives of $E(d,P,L)$ by $P$ are given as 
\begin{align}
    \frac{dE}{dP}=&\frac{1}{d_{min}^{(b)}}\cdot {\rm{ln}}\, \frac{Nx}{eP}+\frac{(2+k^{(b)}{\rm{ln}} \, 2)}{d_{min}^{(b)}}-(2-\frac{3}{d_{min}^{(c)}}){\rm{ln}}\,\frac{e(d-P-L)}{N} - \frac{(2+k_{max}^{(c)}{\rm{ln}} \, 2)}{d_{min}^{(c)}} \nonumber \\
    &+1+{\rm{ln}}\,P + 1 -{\rm{ln}}\,\frac{N}{\mu} <0 , \nonumber \\
    \frac{d^{2}E}{dP^{2}}=&-\frac{1}{d_{min}^{(b)}P}+(2-\frac{3}{d_{min}^{(c)}})\cdot \frac{1}{d-P-L}+\frac{1}{P} > 0. \nonumber
\end{align}
Since the first derivative over $P$ is negative and the second derivative is positive, $E(d,P,L)$ is upper bounded by $$\underset{P \to 0+}{\lim}E(d,P,L)=(2-\frac{3}{d_{min}^{(c)}})(d-L){\rm{ln}}\,\frac{d-L}{N}+\frac{3(2+k_{max}^{(c)})}{d_{min}^{(c)}}\cdot (d-L)+\frac{2(2+k_{max}^{(c)}{\rm{ln}}\,2)}{d_{min}^{(c)}}L.$$ Since the resulting upper bound of $E(d,L)$ is the same as~(37) in \cite{Tdmin_GLDPC}, the rest of the proof is the same as that in \cite{Tdmin_GLDPC} and thus the proposed constraint guarantees the existence of typical minimum distance of the proposed protograph-based PD-GLDPC code.

\section*{Acknowledgment}

The authors would like to thank...

\ifCLASSOPTIONcaptionsoff
  \newpage
\fi

\end{document}